%% file: main_arxiv.tex
\documentclass[aps, prx, amssymb, longbibliography, twocolumn, superscriptaddress, 10pt, nofootinbib]{revtex4-2}

\usepackage[utf8]{inputenc}
\usepackage{booktabs}
\usepackage{graphicx} 
\usepackage{dcolumn} 
\usepackage{bm} 
\usepackage{braket}

\usepackage{tabularx}

\usepackage{listings}
\usepackage{xcolor}
\usepackage{tikz}
\usetikzlibrary{quantikz2}
\usepackage{hyperref}
\hypersetup{                    
	colorlinks,
	citecolor=blue,
	filecolor=blue,
	linkcolor=blue,
	urlcolor=blue
}

\usepackage{orcidlink}

\begin{document}

\title{Extracting the spin excitation spectrum of a fermionic system using a quantum processor}
	
\author{Lucia Vilchez-Estevez \orcidlink{0009-0006-1997-0818}}
\email{lucia.vilchezestevez@physics.ox.ac.uk}
\affiliation{Clarendon Laboratory, University of Oxford, Parks Road, Oxford, OX1 3PU, United Kingdom}
\affiliation{Phasecraft Ltd., Colston Street, Bristol, BS1 4XE, United Kingdom}
\author{Raul A. Santos \orcidlink{0000-0003-1161-1328}}
\affiliation{Phasecraft Ltd., Charlotte Street, London, W1T 4PW, United Kingdom}
\author{Sabrina Yue Wang \orcidlink{0000-0001-8431-6683}}
\affiliation{Phasecraft Ltd., Colston Street, Bristol, BS1 4XE, United Kingdom}
\author{Filippo Maria Gambetta \orcidlink{0000-0002-9182-1415}}
\email{filippo@phasecraft.io}
\affiliation{Phasecraft Ltd., Colston Street, Bristol, BS1 4XE, United Kingdom}

\date{\today}
	
\begin{abstract}
    Understanding low-energy excitations in fermionic systems is crucial for their characterization. They determine the response of the system to external weak perturbations, its dynamical correlation functions, and provide mechanisms for the emergence of exotic phases of matter. In this work, we study the spin excitation spectra of the 1D Fermi-Hubbard model using a digital quantum processor. Introducing a protocol that is naturally suited for simulation on quantum computers, we recover the retarded spin Green's function from the time evolution of simple observables after a specific quantum quench. We exploit the robustness of the protocol to perturbations of the initial state to minimize the quantum resources required for the initial state preparation, and to allocate the majority of them to a Trotterized time-dynamics simulation. This, combined with the intrinsic resilience to noise of the protocol, allows us to accurately reconstruct the spin excitation spectrum for large instances of the 1D Fermi-Hubbard model without making use of expensive error mitigation techniques, using up to 30 qubits of an IBM Heron r2 device.  
\end{abstract}

\maketitle

\section{Introduction}

Many properties of quantum many-body systems can be obtained by studying their response to an external perturbation~\cite{Nozieres_1999, Giuliani_2005, Martin_2016}. In the limit of weak perturbations around the ground state, the system's response is completely determined by its equilibrium properties through its response functions. The latter describe how the system absorbs (emits) energy from (into) an external probe and can be related to some of the system's correlation functions, the so-called dynamical structure factors, by the fluctuation-dissipation theorem~\cite{Kubo_1966}. The dynamical structure factors are primarily determined by the excitation spectrum of the modes that couple to the perturbation. Remarkably, they can be accessed experimentally by studying the doubly-differential cross-section of inelastic scattering experiments~\cite{Sturm_1993}, such as inelastic neutron scattering (INS)~\cite{Zaliznyak_2015}, resonant inelastic X-ray spectroscopy (RIXS)~\cite{Ament_2011}, and electron energy-loss spectroscopy (EELS)~\cite{Egerton_2009}. The direct connection with experimental probes makes dynamical structure factors an ideal tool to test microscopic models of materials and to compare their predictions with experimental data. 
    
In general, the unbiased computation of dynamical structure factors of many-body quantum systems is a challenging task for state-of-the-art numerical techniques, which are limited to small system sizes, short-time evolutions, or low-energy sectors~\cite{Paeckel_2019, Kuhner_1999, Nocera_2016, Li_2021, Huang_2021, Simkovic_2022}. On the other hand, both analogue and digital quantum simulators provide a natural platform to investigate the dynamics of quantum systems~\cite{Smith_2019, Arute_2020, Sun2023ProbingSimulators, Vijayan_2020, Tarruell_2018, Bohrdt_2021, GuardadoSanchez_2021, Ji_2021, Sopena_2021, Urbanek_2021, Vovrosh_2021, Vovrosh_2021b, Senaratne_2022, Miessen_2023, Hemery_2023, Gomes_2023, Rosenberg_2024, Cochran_2024, Bespalova_2024}. In spite of that, the computation of unequal-time correlators required to obtain the dynamical structure factors remains a hard task as it requires controlled evolutions, auxiliary qubits, or extra measurements~\cite{Somma_2002, Pedernales_2014, Endo_2020}. In this context, protocols capable of extracting spectral properties of low-lying excitations by inspecting the dynamics of quantum systems would be extremely valuable~\cite{Jurcevic_2015, Menu_2018, Baez20, Kocku_2024}. Among them, the recently introduced quench spectroscopy protocol allows one to compute an approximation of the dynamical structure factors from the dynamics of single-particle observables or equal-time correlators~\cite{Villa19, Villa20, Villa_2021, Despres_2024}. It has already been successfully implemented to compute spectral properties of spin and Bose-Hubbard models in both analogue and digital quantum simulators~\cite{Sun2023ProbingSimulators, Chen_2024}. Here, one perturbs the system from its equilibrium state via a local (or global) quantum quench~\cite{Polkovnikov_2011}, evolves it for a time $t$ according to the Hamiltonian under investigation, and measures a carefully selected local observable (or equal-time correlator) which depends on the specific excitation spectrum one would like to probe. The space-time Fourier transform of such a signal is then used to compute an approximation of the corresponding dynamical structure factor (denominated quench spectral function) and to access the spectral properties of the system's low-lying excitations.
	
In this work, we adapt the quench spectroscopy protocol to study the spin excitation properties of fermionic systems. We prove that, by carefully choosing the quench operator, this approach gives access to the exact retarded spin correlation function and dynamical spin structure factor on systems where the initial state does not break the spin reflection symmetry. We then apply this approach to study the spin excitation spectrum of the one-dimensional (1D) Fermi-Hubbard model on an IBM digital quantum computer.  
    
Despite their simplicity, Fermi-Hubbard models are iconic and well-studied prototypes of strongly correlated electronic systems and have been identified as an ideal candidate for simulations on near-term quantum computers~\cite{Wecker_2015a, Wecker_2015b, jiang2018quantum, Montanaro_2020, Arute_2020, Cade_2020, Clinton_2021, Stanisic_2022, Dalzell_2023, Hemery_2023, Nigmatullin_2024, Bespalova_2024, AuYeung_2024}. For example, it is widely believed that they can shed light on the mechanism behind high-temperature superconductivity in cuprates~\cite{Imada_1998, Lee_2006, Keimer_2015, LeBlanc_2015, Fradkin_2015, Jiang_2019, Arovas_2022, Qin_2022}. Exploring the behavior of elementary excitations across different regions of the parameter space of Fermi-Hubbard models can offer critical insights into this phenomenon. In particular, the study of dynamical (density and spin) structure factors has proven to be a powerful tool for characterizing these excitations and advancing the  understanding of the underlying physics~\cite{Zaliznyak_2004, Eschrig_2006, Walters_2009, Jia_2014, Li_2021}. In the 1D case, it is well known that low-energy excitations of the Fermi-Hubbard model can be described in terms of (anti-)holons and spinons which, due to interplay between dimensionality and interactions, can give rise to peculiar phenomena, such as spin-charge separation~\cite{Giamarchi_2003, Solyom10, Arute_2020}. In this work, we focus mainly on spin excitations, which are of great relevance to high-temperature superconductivity~\cite{Dagotto_1999,Orenstein_2000, Maekawa_2001, Dong_2022} and the search of quantum spin liquids~\cite{Wen_2007, Luo_2018, Lancaster_2023, Gohlke_2024}. Their spectrum has been measured in INS and RIXS experiments on quasi-1D chains whose physics is well captured by a 1D Fermi-Hubbard model, such as Sr$_2$CuO$_3$~\cite{Zaliznyak_2004, Walters_2009, Schlappa_2012, Schlappa_2018}.
	
When implementing a time-dynamics simulation on a near-term (i.e., NISQ) quantum computer, one has to consider that, due to the presence of noise, the computational resources (i.e., number of qubits, number of gates, and circuit depth) are limited and must be carefully optimized~\cite{Preskill_2018, Bharti_2022}. A generic quench spectroscopy experiment consists of four steps: initial state preparation, quench, time evolution, and measurement. Resource-wise, given a fixed number of qubits and a quantum architecture where measurements are fast, initial state preparation and time evolution are the most expensive steps. In particular, one of the main challenges for these kinds of experiment is that, usually, the preparation of a non-trivial initial state (such as the ground state of system) already requires a significant amount of resources~\cite{Arute_2020}. In this work, we exploit the remarkable robustness of the quench spectroscopy protocol with respect to perturbations of the initial state~\cite{Villa19, Villa20} to minimize the amount of resources allocated to the initial state preparation. This robustness comes in two flavours: resilience to low levels of hardware noise and to small perturbations of the initial state of the quench spectroscopy protocol compared to the energy gap in the spectrum of the spin excitations. We show that applying the quench to the ground state of a free-fermion model allows us to recover an excellent approximation to the spin excitation spectrum of the Fermi-Hubbard model up to intermediate interaction strengths.
Although this state can be efficiently prepared using Givens rotations~\cite{jiang2018quantum}, it still requires a circuit whose depth scales linearly with the system size. Hence, we further reduce the resources required for preparing the initial state by introducing the Dense Givens Approximate (DGA) state preparation scheme, in which a size-independent shallow ansatz consisting of layers of parameterized Givens rotations is used to prepare an approximation of the free-fermion ground state. We show that the quench spectral function computed using this initial state is in good agreement with the original one. This allows us to devote the majority of the computational resources to the Hamiltonian simulation part of the protocol, which is performed using a first-order Trotter formula. Here, we studied different encodings of the fermionic degrees of freedom to further reduce the depth of a single layer of the Trotter circuits to simulate the dynamics of the system. All of these optimizations allow us to perform a time-dynamics simulation on an IBM quantum computer of Fermi-Hubbard model instances up to size $L = 15$ and recover all the features of the two-spinon continuum spectrum. 
		
The paper is organized as follows. In Section \ref{sec:protocol} we discuss the unitary quench spectroscopy protocol and its application to fermionic systems. In Section \ref{sec:classical_simulations}, we describe the details of the protocol for extracting spectral properties of the spin excitations of the 1D Fermi-Hubbard model and show results obtained by tensor network simulations. As the evolution of this 1D system is tractable using classical numerical techniques, we use these results to study the best parameters to perform the quantum experiment. Here, we also discuss the resilience of the protocol to different perturbations and errors. In Section \ref{sec:implem} we discuss the details of the quantum circuit implementation of the quench spectroscopy protocol. In Section \ref{sec:approx_ff} we introduce the DGA scheme to prepare the approximate free-fermion ground state using a shallow circuit. In Section \ref{sec:results} we first discuss the resources required to implement the protocol on IBM quantum hardware, and then we show the results obtained from experiments on the state-of-the-art IBM Heron r2 device ibm\_fez.

\section{Unitary quench spectroscopy protocol}
\label{sec:protocol}

In a typical linear response theory setting, an external field (the probe) couples linearly with a system's observable $B$ and induces dynamics into another system's observable $A$. The response of the system to such a perturbation is characterized by the following retarded Green's function (also known as the retarded linear response function)~\cite{Giuliani_2005, Martin_2016}
\begin{equation}
    \label{eq:greens_function}
    G_{AB}(t) := -i \braket{\Psi|[A(t),B]|\Psi} \quad \text{with }t\geq0,
\end{equation}
where $\ket{\Psi}$ is usually the ground state of the system, $[A,B]:=AB-BA$ is the commutator, and $A(t)$ is the Heisenberg evolved operator with respect to the system's Hamiltonian $H$, $A(t):=e^{-iHt}Ae^{iHt}$. Here, we have set $\hbar = 1$ and, for the sake of simplicity, we have omitted any spatial dependence. For the important case $B=A^{\dagger}$, the fluctuation-dissipation theorem links the imaginary part of the Fourier transform of $G_{AA^\dagger}(t)$ to the system's dynamical structure factor $\mathcal{S}_{AA^\dagger}(\omega)$, which is defined as the Fourier transform of the correlation function $\mathcal{S}_{AA^\dagger}(t): = \braket{\Psi|A(t)A^\dagger|\Psi}$ and can be obtained from inelastic scattering experiments~\cite{Sturm_1993}. At zero-temperature, the fluctuation-dissipation theorem reads~\cite{Giuliani_2005}
\begin{equation}
    \label{eq:FD}
    \mathrm{Im}G_{AA^\dagger}(\omega) = -2\pi \mathrm{sign}(\omega)\mathcal{S}_{AA^\dagger}(\omega).
\end{equation} 
In particular, one finds
\begin{equation}
    \label{eq:ImG}
    \mathrm{Im}G_{AA^\dagger}(\omega)=-\pi\sum_{n}|A_{0n}|^2\left[\delta(\omega-E_n)-\delta(\omega+E_n)\right],
\end{equation}
where $A_{0n} = \braket{\Psi|A|E_n}$, with $\ket{E_n}$ eigenstate of $H$ with eigenvalue $E_n$. Note that the ground-state energy $E_0$ is set to zero. Therefore, spectral information about the low-lying excitations of the system that couple to the operator $A$ can be obtained by looking at $\mathrm{Im} G_{AA^\dagger}(\omega)$ for $\omega>0$~\cite{Giuliani_2005, Martin_2016}.

The operator $B$ in Eq.~\eqref{eq:greens_function} has an expansion in terms of a string of Pauli operators (i.e., $B:=\sum_j \alpha^B_j P_j$). So, without loss of generality, we can focus on $G_a(t):=i\langle \Psi|[P_a,A(t)]|\Psi\rangle$ for a Pauli string $P_a$. As Pauli operators square to the identity, it follows that
\begin{equation}
G_a(t)=\sum_{\sigma=\pm}\sigma\langle\Psi|e^{i\sigma\frac{\pi}{4}P_a}A(t)e^{-i\sigma\frac{\pi}{4}P_a}|\Psi\rangle.
\end{equation}
The interpretation of this relation is simple: It corresponds to the dynamics of the operator $A$ after a local quench over the state $|\Psi\rangle$ with the unitary operator $e^{i\sigma\frac{\pi}{4}P_a}$. The retarded Green's function $G_{AB}$ can therefore be computed by measuring the time-evolved operator $A$ on the quenched state. This represents a generalization of the local quench spectroscopy protocol introduced in Refs.~\cite{Villa19,Villa20}. The latter allows one to recover the spectrum of low-lying excitations of many-body quantum systems by studying the non-equilibrium dynamics of single-particle observables after a quantum quench and is therefore perfectly suited for current quantum simulators~\cite{Baez20, Roushan17, Lee21, Sun2023ProbingSimulators, Chen_2024}. 
The number of measurements in this protocol then depends on the number of terms that appear in the expansion of the operator $A$. We can optimize this further by considering the particular structure of the problem. To this end, we specialise on spin observables in fermionic systems. The spin operators are defined as bilinears in the fermion creation (annihilation) operators $c^\dagger_{i\sigma} (c_{i\sigma})$ at site $i$ and spin $\sigma$,
\begin{align}
\hat{S}_i^\alpha:=\sum_{\sigma\sigma'}c_{i,\sigma}^\dagger S^\alpha_{\sigma\sigma'}c_{i,\sigma'},
\end{align}
where $S^\alpha$ is the $\alpha = x, y, z$ Pauli matrix. The retarded spin Green's function is then defined as~\footnote{In the literature this function is also known as spin response function and is usually denoted by $\chi_{jk}(t)$~\cite{Simkovic_2022}.}
\begin{equation}
    \label{eq:spin_greens_func}
    G_{jk}^{x}(t) :=-i\langle \Psi|[\hat{S}_{k}^{x}(t),\hat{S}_{j}^{x}]|\Psi\rangle  \quad \text{with }t\geq0.
\end{equation}
Using the fermionic anticommutation relations, it is easy to prove that $(\hat{S}_{i}^{x})^{2k}=(\hat{S}_{i}^{x})^{2}$ and $(\hat{S}_{i}^{x})^{2k+1}=\hat{S}_{i}^{x}$. From these relations it follows that 
\begin{equation}
\label{eq:exp}
e^{i\theta \hat{S}_{j}^{x}}=1+(\hat{S}_{j}^{x})^{2}(\cos\theta-1)+i\sin\theta \hat{S}_{j}^{x}.
\end{equation}
Using this identity and defining the state 
\begin{align}
    \label{eq:exp_2}
    \ket{\Psi_{\theta}^j}&:=e^{i\theta \hat{S}_{j}^{x}}\ket{\Psi}\nonumber\\
    &= \left[1+(\hat{S}_{j}^{x})^{2}(\cos\theta-1)+i\sin\theta \hat{S}_{j}^{x}\right]\ket{\Psi}, 
\end{align}
we obtain
\begin{align}
    \label{eq:GF_quench}
    \bra{\Psi_{\theta}^j}\hat{S}^x_k(t)
    \ket{\Psi_{\theta}^j} &= i\sin\theta\Big[\bra{\Psi} [\hat{S}^x_j,\hat{S}^x_k(t)]\ket{\Psi} \nonumber\\ 
    &-(\cos\theta-1) \bra{\widetilde{\Psi}^x_j} [\hat{S}^x_j,\hat{S}^x_k(t)] \ket{\widetilde{\Psi}^x_j}\Big] \\
    &= \sin\theta\left[G^x_{jk}(t) - (\cos\theta-1)\ \widetilde{G}^x_{jk}(t)\right], \nonumber
\end{align}
where we defined $\ket{\widetilde{\Psi}^x_j}:=\hat{S}^x_j\ket{\Psi}$ and $\widetilde{G}^x_{jk}(t) := -i\bra{\widetilde{\Psi}^x_j}[\hat{S}_{k}^{x}(t),\hat{S}_{j}^{x}]\ket{\widetilde{\Psi}^x_j}$, for $t>0$.
In deriving Eq.~\eqref{eq:GF_quench}, we used that for any state $\ket{\Psi}$ that is an eigenstate of the total spin projection operator $\hat{S}^z_{\rm tot} = \sum_{j}\hat{S}^z_j$ and a Hamiltonian $H$ such that $[H, \hat{S}^z_{\rm tot}] = 0$, the expectation of an odd number of $\hat{S}_{j}^{x}$ operators is zero, as each of them raise or lower the total spin projection by one. Note that the 1D Fermi-Hubbard model we consider in this work satisfies both these conditions. For small values of $\theta$, Eq.~\eqref{eq:GF_quench} becomes
\begin{equation}
\bra{\Psi_{\theta}^j}\hat{S}^x_k(t)\ket{\Psi_{\theta}^j} \approx \theta G^{x}_{jk}(t) - \theta^3\widetilde{G}^x_{jk}(t).
\end{equation}
Therefore, the exact retarded spin Green's function of a fermionic system, $G^{x}_{jk}(t)$, can be obtained from the dynamics of $\hat{S}^x_k$ with arbitrary precision in the limit $\theta \rightarrow 0$ (or $\theta \rightarrow \pi$). For this choice of $\theta$, however, the signal is suppressed and this makes it hard to retrieve it in the presence of noise. In general, one has therefore to work in a regime in which the contribution of $\widetilde{G}^x_{jk}(t)$ to the expectation value of $\hat{S}^x_k(t)$ is non-negligible.

Recalling Eq.~\eqref{eq:ImG}, the spectral properties of low-lying spin excitations can be extracted from the imaginary part of the Fourier transform of $G^x_{jk}(t)$ (which, by virtue of the fluctuation-dissipation theorem of Eq.~\eqref{eq:FD}, is also connected to the dynamical spin structure factor). When $\widetilde{G}^x_{jk}(t)\neq 0$, the (imaginary part of the) space-time Fourier transform of $\bra{\Psi_{\theta}^j}\hat{S}^x_k(t)\ket{\Psi_{\theta}^j}$ only gives access to an approximation of the dynamical structure factor. In the language of local quench spectroscopy introduced in Ref.~\cite{Villa20}, such a quantity is known as the quench spectral function (QSF), and requires a careful choice of the quench operator and of the observable to measure in order to avoid contributions from sectors of the system's Hilbert space with a different number of excitations. For a fermionic system, however, the quenched state $\ket{\Psi^j_\theta}$ of Eq.~\eqref{eq:exp_2} is a low-energy state with support only in the sectors of the system's Hilbert space with total spin projection eigenvalues $S^z_{\mathrm{tot}}$ and $S^z_{\mathrm{tot}} \pm 1$, with $\hat{S}^z_{\mathrm{tot}}\ket{\Psi} = S^z_{\mathrm{tot}} \ket{\Psi}$. Indeed, since $(\hat{S}_{j}^{x})^2 = n_{j, \uparrow} + n_{j, \downarrow} - 2n_{j, \uparrow}n_{j, \downarrow}$, with $n_{i,\sigma}=c_{i,\sigma}^{\dagger}c_{i,\sigma}$ the density operator for the spin $\sigma \in \{\uparrow, \downarrow \}$ on site $i$, the action of $(\hat{S}^x_j)^2$ on the state $\ket{\Psi}$ does not change its total spin projection. Therefore, by virtue of the spin-charge separation of the 1D Fermi-Hubbard model~\cite{Solyom10}, the dynamics of $\bra{\Psi_{\theta}^j}\hat{S}^x_k(t)\ket{\Psi_{\theta}^j}$ is dominated by two-spinon excitations and we expect the QSF to probe almost exactly the same two-spinon continuum as the dynamical spin structure factor (although their spectral weight may differ slightly)~\cite{Villa20} (see Fig.~\ref{fig:comparative_panel}(a-j) in Section~\ref{sec:results_sims} below for a numerical comparison between these two quantities).

\begin{figure}
    \centering
    \includegraphics[width=0.9\columnwidth]{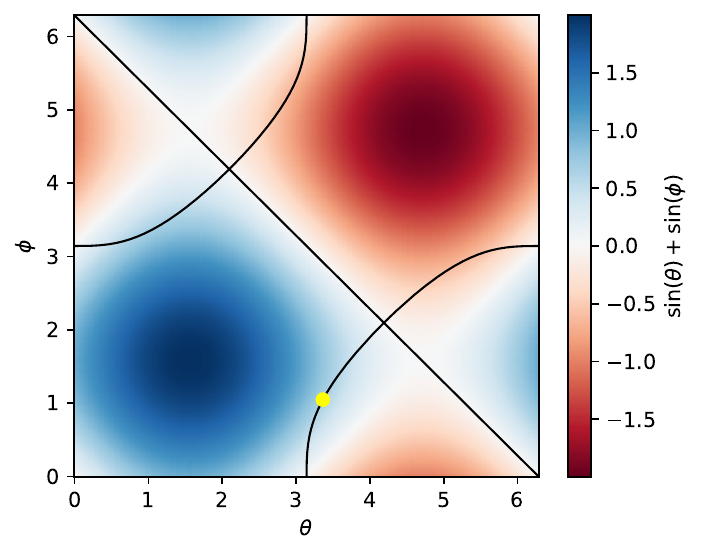}
    \caption{Density plot of the coefficient of $G^x_{jk}(t)$ in the double quench spectroscopy protocol of Eq.~\eqref{eq:GF_double_quench_spin}, $\sin\theta+\sin\phi$. The black contours denote the region of the $\theta-\phi$ plane in which Eq.~\eqref{eq:cos_condition} is satisfied. The yellow dot, corresponding to $(\theta, \phi) \approx (1.07\pi, \pi/3)$, represent an almost optimal pair of angles for the protocol.}
    \label{fig:double_quench_conditions}
\end{figure}

We now discuss how our protocol can be adapted to access the exact retarded spin Green's function (and, therefore, the exact dynamical spin structure factor). First, this occurs for any state $\ket{\Psi}$ such that $(\hat{S}^x_j)^2\ket{\Psi} = \lambda\ket{\Psi}$, $\forall j$, $\widetilde{G}^x_{jk}(t) = -\lambda G^x_{jk}(t)$. In this case, Eq.~\eqref{eq:GF_quench} becomes
\begin{align}
    \bra{\Psi_{\theta}^j}\hat{S}^x_k(t)
    \ket{\Psi_{\theta}^j} &= \sin\theta\left[1+\lambda(\cos\theta-1)\right]G^{x}_{jk}(t).
\end{align}
Since $(\hat{S}_{j}^{x})^2 = n_{j, \uparrow} + n_{j, \downarrow} - 2n_{j, \uparrow}n_{j, \downarrow}$, the condition above is satisfied only for states $\ket{\Psi}$ in which every site $j$ has a well defined occupation (such as, for instance, the N\'{e}el state~\cite{Martin_2016}). For more general states, the exact retarded spin Green's function of a fermionic system can always be obtained from Eq.~\eqref{eq:GF_quench} by combining two quench experiments from the states $\ket{\Psi^j_{\theta}}$ and $\ket{\Psi^j_{\phi}}$, with $\theta$ and $\phi$ such that
\begin{subequations}
\label{eq:double_quench_conditions}
\begin{align}
    \sin\theta + \sin\phi &\neq 0, \label{eq:sin_condition}\\
    \sin\theta(\cos\theta-1) + \sin\phi(\cos\phi-1) &= 0. \label{eq:cos_condition}
\end{align}
\end{subequations}
As shown in Fig.~\ref{fig:double_quench_conditions}, there exist infinitely many pairs $(\theta,\phi)$ satisfying Eqs.~\eqref{eq:double_quench_conditions}. For each one of such pairs, the exact retarded spin Green's function can be obtained via
\begin{align}
\label{eq:GF_double_quench_spin}
&\bra{\Psi_{\theta}^j}\hat{S}^x_k(t)\ket{\Psi_{\theta}^j} + \bra{\Psi_{\phi}^j}\hat{S}^x_k(t)\ket{\Psi_{\phi}^j}\nonumber\\ &= (\sin\theta + \sin\phi)G^{x}_{jk}(t).
\end{align}
The protocol above can be viewed as a generalization to fermionic systems of the many-body Ramsey interferometry technique introduced for spin systems~\cite{Knap_2013_probing, Schuckert_2020_probing, Baez20}.
To get a good signal, it is important to choose a pair $(\theta, \phi)$ that maximizes the coefficient of $G^x_{jk}(t)$ in Eq.~\eqref{eq:GF_double_quench_spin}, $\sin\theta + \sin\phi$. An example of such a pair is represented by the yellow dot in Fig.~\ref{fig:double_quench_conditions}, corresponding to $(\theta, \phi) \approx (1.07\pi, \pi/3)$, for which one finds $\sin\theta + \sin\phi \approx 0.65 $.
While in this work we focus on the spin response, in Appendix~\ref{app:charge_response} we show that a similar approach allows one to probe the exact charge response as well.

In what follows, we apply this protocol to the 1D Fermi-Hubbard model on $L$ sites with open boundary conditions and lattice spacing equal to 1, with Hamiltonian
\begin{align}
    \label{eq:FH}
    H_{\rm FH}&:=-J\sum_{i = 0}^{L-2}\sum_{\sigma}(c_{i,\sigma}^\dagger c_{i+1,\sigma}+\text{h.c.})+U\sum_{i=0}^{L-1} n_{i,\uparrow}n_{i,\downarrow}\nonumber\\
    &:=H_0 + H_{\rm{int}},
\end{align} 
where $J$ is the hopping strength and $U$ is the on-site interaction parameter (or Coulomb potential).
In this work, we will fix $J=1$ and $U/J=3$ unless stated otherwise. All energies (times) will be measured in units of $J$ ($1/J$). Low-energy excitations of the Fermi-Hubbard model can be generated through multiple mechanisms. When the total number of electrons per spin sector is conserved, electron-hole pairs can be formed. Charged excitations arise by either adding or removing particles with spin-up or spin-down from the ground state. Here, we focus on spin excitations, which are produced by flipping the spin of an electron and can be described in terms of a pair of spinons~\cite{Solyom10}. Since the ground state of the 1D Fermi-Hubbard model has definite total spin and is symmetric under spin reflection, we can access its QSF using Eq.~\eqref{eq:GF_quench}. As we will show in Section~\ref{sec:classical_simulations} (see Fig.~\ref{fig:comparative_panel}(a-j)), the agreement between the QSF obtained from a single quench via Eq.~\eqref{eq:GF_quench} with $\theta = \pi/4$ and the exact dynamical spin structure factor is excellent. Therefore, in this work we have not used the double quench experiment approach described in Eqs.~\eqref{eq:double_quench_conditions} as it requires additional computational overhead. Furthermore, we will show in Section~\ref{sec:classical_simulations}, for the values of the system parameters we consider here one can recover the spectroscopic features of the low-energy excitations of the Fermi-Hubbard model by applying the quench to the ground state of a free-fermion model or even to a shallow circuit approximation of the latter (see Section~\ref{sec:approx_ff}). Both states have definite total spin and are symmetric under spin reflection, which means that Eq.~\eqref{eq:GF_quench} still holds. 

\section{Classical simulation, initial state preparation, and Trotterization requirements}
\label{sec:classical_simulations}
The protocol outlined above consists of four main stages. First, initialize the system to the ground state (or to a quantum state that closely approximates it). Second, apply a local quench to introduce a low-energy excitation, whose spectrum is the focus of our study. Third, the evolve in time the perturbed state over a specified duration $t$. Finally, measure the expectation value of the spin operator $\hat{S}_j^x$ to analyze the resulting dynamics.

As discussed in the previous section, the quench operator is given by
\begin{equation} 
    \label{eq:quench_op_unitary}    
    \hat{Q}_j := e^{i \frac{\pi}{4} \hat{S}^x_j},
\end{equation}
To minimize the effect of the open boundary conditions, we apply the quench operator to the central site $j = \lfloor L/2\rfloor$. After the quench, the state is evolved under the time-evolution operator associated with Eq.~\eqref{eq:FH}. Once the time-evolved state $\ket{\Psi^{j}_{\frac{\pi}{4}}(t)}=e^{-iHt}\ket{\Psi^{j}_{\frac{\pi}{4}}}$ is constructed, we measure the expectation value of the $\hat{S}_i^x$ operator for each site $i \in \{0,..., L-1\}$. 

\subsection{Classical simulation}
\begin{figure}
    \centering
    \includegraphics[width=\linewidth]{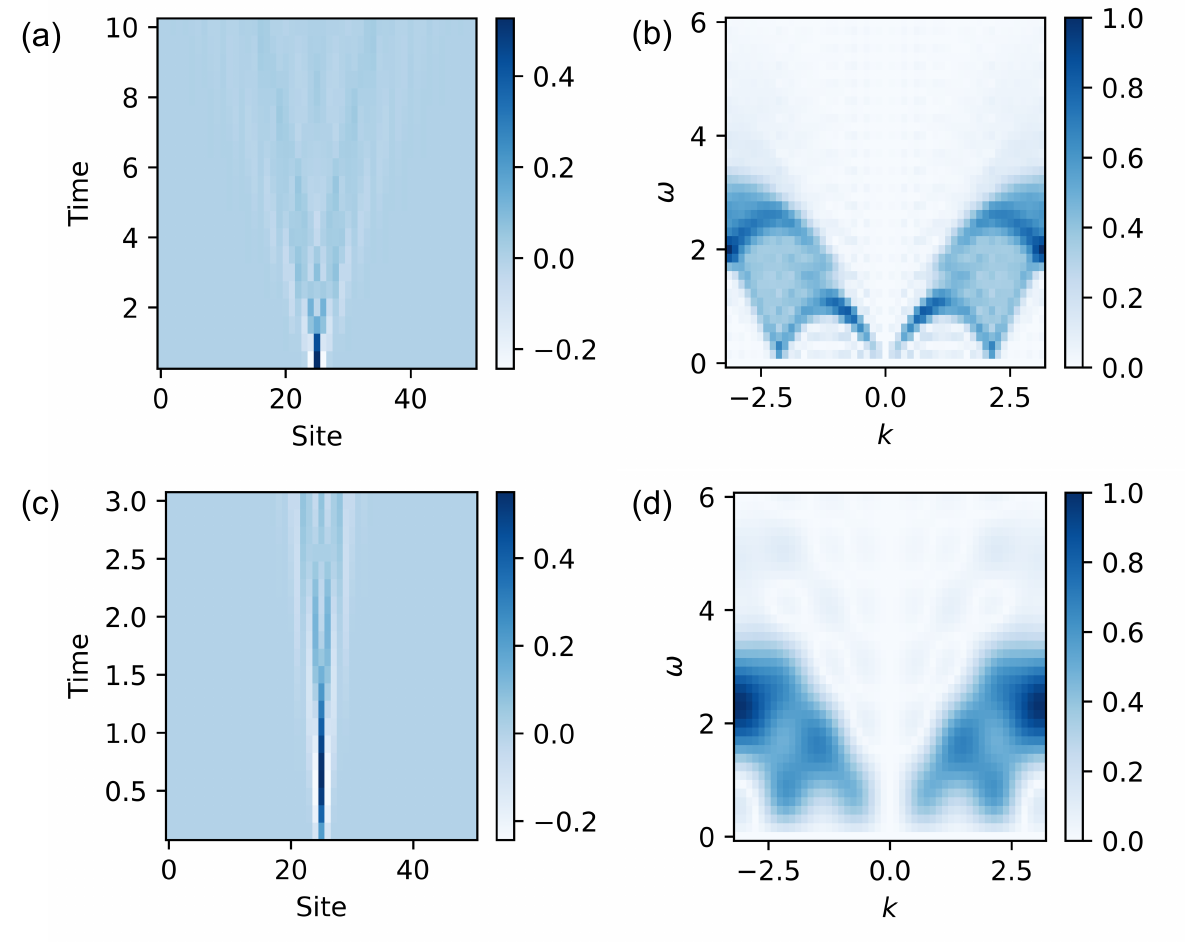}
    \caption{TDVP simulation of the quench protocol for the 1D Fermi-Hubbbard model with $L=51$ sites. (a),(c) Dynamics of the observable $\braket{\hat{S}^x_i(t)}$ for $U=3$ and filling $n_e=2/3$ for total running times $T=10$ and $T=3$, respectively. (b), (d) Normalised absolute value of the corresponding quench spectral functions in the momentum-frequency domain after applying a two-dimensional Fourier transform to the space-time domain data. The maximum bond dimension for the TDVP simulation is $\chi = 128$.
    }
    \label{fig:classical_simulation}
\end{figure}

An example of the protocol for a system with $L=51$ sites and long-time evolution is shown in Fig.~\ref{fig:classical_simulation}. The simulation has been performed using the implementation of the time-dependent variational principle (TDVP) algorithm~\cite{Haegeman11,Haegeman16} provided by TenPy~\cite{tenpy}. Here, we fixed $U=3$ and filling fraction $n_e:=N_e/L = 2/3$, with $N_e =\sum_{i, \sigma} \braket{n_{i,\sigma}}_0$ the total number of electrons in the initial state $\ket{\Psi}$. Here, $\braket{...}_0$ denotes the expectation value on the Fermi-Hubbard model ground state, $\ket{\Psi}$. Once $\braket{\hat{\boldsymbol{S}^x}(t)} = (\braket{\hat{S}^x_0(t)}, ..., \braket{\hat{S}^x_{L-1}(t)})$ (with $\braket{...}$ the expectation value on $\ket{\Psi^j_{\frac{\pi}{4}}}$) is obtained for all the time steps $t_m$ in the time evolution (with $m=1,...,M$ and $t_{M-1} = T$), the spin excitation spectrum can be resolved by taking a space-time Fourier transform of $\braket{\hat{\boldsymbol{S}^x}(t)}$ and examining the data in the momentum-frequency domain. This gives access to the QSF, as discussed above. In this work, all Fourier transforms are computed using the zero-padding technique with respect to time, whereby a number $N'_\mathrm{samples} - N_\mathrm{samples}$ of zeros is appended to the original time signal. In particular, $N'_\mathrm{samples}$ is chosen in such a way that $2N_\mathrm{samples}$ frequencies are present in the frequency interval $\omega \in [0, 6J]$. See Appendix~\ref{app:fourier} for more details.

The presence of noise in current near-term quantum computers limits the maximum evolution time that can be reached on a quantum computer and makes long-time evolutions impractical due to the accumulation of errors. Therefore, we aim to determine the shortest evolution time that still allows us to recover the main properties of the spin excitation spectrum. For example, one peculiar feature of the two-spinon excitation spectrum is the dome-shaped structure with zero-frequency cusps at $k=0$ and $ k = \pm 2k_F = \pm \pi n_e$, which can be used in experiments to derive information about the doping of the system. Here, $k_F:= \pi n_e/2$ is the Fermi momentum.
From Fig.~\ref{fig:classical_simulation}, we observe that although the resolution decreases and the spectral lines broaden, a shorter evolution time of $T=3$ is sufficient to recover the spectrum. Shorter times result in greater uncertainty in features such as the precise location of the zero-frequency cusps. However, we argue that this level of accuracy is often sufficient and can significantly reduce computational resource requirements. Furthermore, for smaller systems, the resolution is inherently limited by the system size, meaning that increasing the total evolution time does not necessarily lead to an improved spectrum compared with the thermodynamic limit. 

\subsection{Initial state preparation and free-fermion ground state}
To compute the QSF using the protocol introduced in Section~\ref{sec:protocol}, the quench operator $\hat{Q}_j$ should be applied to the ground state $\ket{\Psi}$ of the 1D Fermi-Hubbard model. However, preparing the ground state of interacting many-body systems on a quantum simulator is a challenging problem on its own~\cite{Schuch_2009, Gharibian_2015} and, in general, costly in terms of circuit implementation~\cite{Peruzzo_2014, McClean_2016, Cade_2020, Cai_2020}. On the other hand, as discussed for the local quench spectroscopy protocol in Ref.~\cite{Villa19}, the spectrum of a given low-energy excitation can be reconstructed by computing the QSF on a different state $\ket{\Psi'}$ such that $\hat{Q}_j\ket{\Psi'} \approx \sum_{\boldsymbol{q}}\alpha_{\boldsymbol{q}} \gamma^{\dagger}_{\boldsymbol{q}}\ket{\Psi}$, with $\gamma_{\boldsymbol{q}}^{\dagger}$ the creation operator in momentum space associated with the excitation under investigation. This holds if the state $\hat{Q}_j\ket{\Psi'}$ contains at most a single excitation. Note that in our case, the low-energy excitations are pairs of spinons. Hence, the above requirement becomes $\hat{Q}_j\ket{\Psi'}\approx\sum_{\boldsymbol{k}}\alpha_{\boldsymbol{k}}\sum_{\boldsymbol{q}}(c^{\dagger}_{\boldsymbol{q},\uparrow}c_{\boldsymbol{k}+\boldsymbol{q},\downarrow}+c^{\dagger}_{\boldsymbol{q},\downarrow}c_{\boldsymbol{k}+\boldsymbol{q},\uparrow})\ket{\Psi}$, where we used the momentum space form of the spin operator $\hat{S}^x_j$ as given in Eq.~\eqref{eq:spin_op_momentum} of Appendix~\ref{app:robustness}.

Here, we examine the performance of the protocol when the quench operator is applied to the ground state of the free-fermion Hamiltonian, $\ket{\psi_{\rm FF}}$. 
This alternative is well-suited for quantum computing, as it can be efficiently implemented using Givens rotations~\cite{jiang2018quantum} and can serve as a good approximation of the Fermi-Hubbard Hamiltonian for small values of $U$. Remarkably, we find that this approximation remains effective even in the presence of strong interactions. 
\begin{figure}
    \centering
    \includegraphics[width=0.92\linewidth]{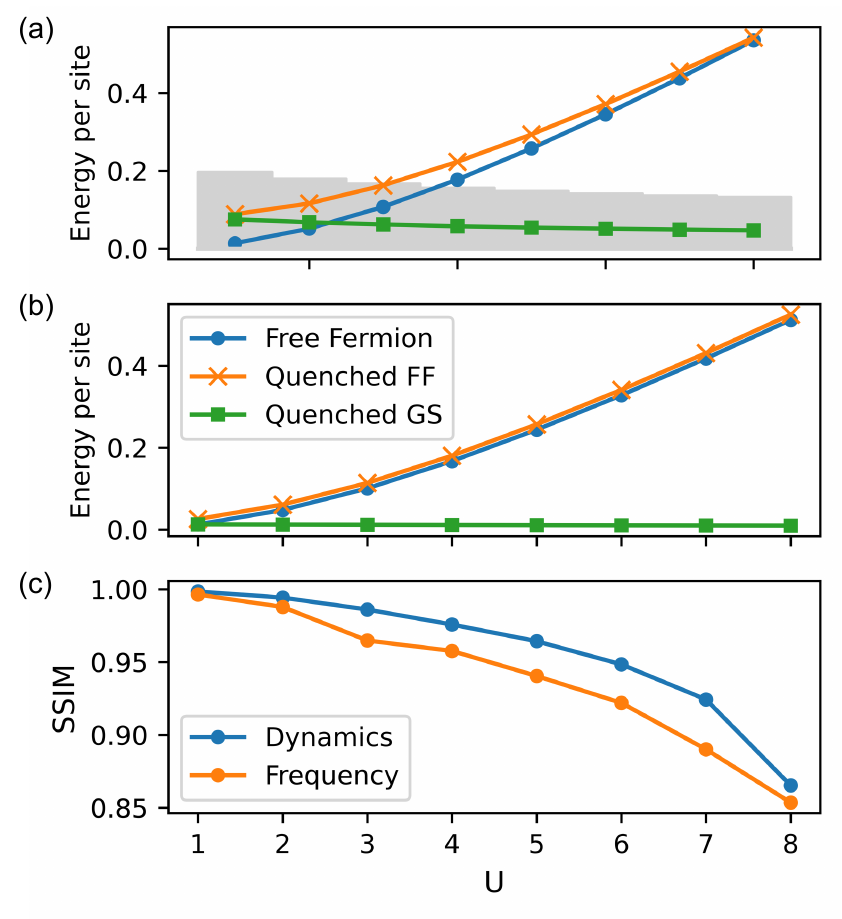}
    \caption{Simulation of the quench spectroscopy protocol for a 1D Fermi-Hubbard model initialized in the free-fermion ground state. (a) Energy per site for different initial states in a system with $L=15$ sites, setting the ground state energy of the Fermi-Hubbard Hamiltonian $H_{\rm FH}$ to zero. The green curve represents the energy of the state obtained by applying the quench to the Fermi-Hubbard ground state, $\hat{Q}_{\lfloor L/2 \rfloor}\ket{\Psi}$, the blue curve shows the expectation value of the Fermi-Hubbard Hamiltonian $H_{\rm FH}$ on the free-fermion ground state $\ket{\psi_{\rm FF}}$, and the orange curve corresponds to the energy of the state obtained by applying the quench to the free-fermion ground state, $\hat{Q}_{\lfloor L/2 \rfloor}\ket{\psi_{\rm FF}}$, all as functions of the on-site interaction $U$. Also shown is the energy range (shaded region) of the subspace spanned by $\ket{\Psi}$ and $\hat{Q}_{j}\ket{\Psi}$ with $j = 0,...,L-1$, i.e., the low-energy subspace of $H_{\rm FH}$ containing at most a single spin excitation. (b) Same as for (a), but for a system with $L=51$ sites. (c) Structural similarity index measure (SSIM) between the space-time signals obtained starting from the Fermi-Hubbard and free-fermion ground state with $L=51$ (blue curve) and their Fourier transforms (yellow curve) for a total evolution time of $T=10$. Values of the SSIM close to 1 indicate that two images are similar, with ${\rm SSIM} = 1$ for identical signals.}
    \label{fig:free_fermion_simulation}
\end{figure}

As we can see in Fig.~\ref{fig:free_fermion_simulation}(a-b), the energy of the Fermi-Hubbard ground state and the expectation value of $H_{\rm FH}$ on the free-fermion ground state diverge fairly quickly as a function of the on-site interaction strength $U$, suggesting that the overlap between the two states decreases with increasing $U$. As can be seen by comparing the blue and orange curves, the expectation value of $H_{\rm FH}$ on the free-fermion ground state is larger than the energy of the state $\hat{Q}_{\lfloor L/2 \rfloor}\ket{\Psi}$ for $U\gtrsim 2$. However, from panel (a) we can see that, for a system with $L=15$, the state $\hat{Q}_{\lfloor L/2 \rfloor}\ket{\psi_{\rm FF}}$ (orange curve) is within the energy range of the subspace of $H_{\rm FH}$ containing at most a single excitation (shaded region) up to $U\lesssim4$. This means that, for larger values of $U$, $\hat{Q}_{\lfloor L/2 \rfloor}\ket{\psi_{\rm FF}}$ may contain more elementary excitations, resulting in a QSF with reduced accuracy.
On the other hand, the structural similarity index measure (SSIM)~\cite{Zhou_2004,Venkataramanan_2021} (see Appendix~\ref{app:ssim} for details) between the spectra obtained through the quench spectroscopy protocol starting from $\ket{\Psi}$ and $\ket{\psi_{\rm FF}}$ remains close to 1 over a considerable range of $U$ even for a large system size (see Fig.~\ref{fig:free_fermion_simulation}(c)). We recall that $-1\leq {\rm SSIM} \leq 1$, with ${\rm SSIM} = 1$ for two identical images. This suggests that additional excitations does not significantly affect the expectation value of $\hat{S}^x_i(t)$, confirming the robustness of the quench spectroscopy protocol against perturbations to the system's ground state~\cite{Villa19, Villa20} and justifying the use of the free-fermion ground state in our simulations and subsequent implementation. For a more in-depth comparison between the QSFs obtained initializing the system in the Fermi-Hubbard model and free-fermion ground state see Fig.~\ref{fig:comparative_panel} in Section~\ref{sec:results}. In Appendix~\ref{app:robustness} we further discuss the robustness of the protocol from the point of view of perturbation theory. We note here that initializing the protocol in the free-fermion ground state instead of the Fermi-Hubbard model's one can be interpreted in two different ways: Either as the exact quenched non-equilibrium spin Green's function, $\langle \Psi_{\rm FF}|[\hat{S}_{k}^{x}(t),\hat{S}_{j}^{x}]|\Psi_{\rm FF}\rangle$, or as an approximation to the equilibrium retarded spin Green's function discussed in Section~\ref{sec:protocol}. In principle, both contain important information about the nature of the excitations of a many-body quantum system. For instance, in the second interpretation, while the approximation is not formally controlled, it can still reveal the breakdown of the quasiparticle picture, as it is always possible to compute the phase space available to the quasiparticle excitations (see, e.g., Appendix~\ref{app:robustness}) and compare it with the signal extracted from a quantum computer. Furthermore, at the end of Section~\ref{sec:DGA_practical} we will discuss how this approximation can be turned into a controlled one at the price of a higher quantum computational cost.

\subsection{Trotterization}
To successfully implement the protocol on a quantum device and obtain the spin excitation spectrum, we need to time-evolve the quenched quantum state. We do this using a product (Trotter) formula, which approximates the time-evolution operator of a Hamiltonian $H = \sum H_i$ as a product of its summands $e^{-iH_it}$~\cite{Child_2019, Childs_2021, Ostmeyer_2023, Dalzell_2023}. The evolution operator is thus decomposed into a product of $N_{\mathrm{Trotter}}$ small time evolution operators, $\mathcal{U}(t)=\exp(-i H t) = \mathcal{U}(\delta t) \dots \mathcal{U}(\delta t)$, with $\delta t = t/N_{\mathrm{Trotter}}$. In the case of the Fermi-Hubbard model and using a first-order Trotter formula, we can further discretize every $\mathcal{U}(\delta t)$ as a product of terms that can be implemented simultaneously on a quantum device, i.e., a product of the interacting part of the Hamiltonian $H_{\rm int}$, the hopping term on even sites $H_0^{\rm even}$, and the hopping term on odd sites $H_0^{\rm odd}$:
\begin{align}
    \mathcal{U}(t) &= (e^{-i\delta tH_0^{\rm even}} e^{-i\delta t H_0^{\rm odd}} e^{-i\delta t H_{\rm int}})^{N_{\rm Trotter}} \nonumber\\ &+O\left(\frac{t^2}{N_{\rm Trotter}}\right).
    \label{eq:Trotter}
\end{align}
Since our main purpose is to find an optimal implementation of the quench spectroscopy protocol in NISQ devices, we will avoid using higher-order Trotter formulas due to higher gate counts and deeper circuits~\cite{Ostmeyer_2023}. Hence, in the following, we will only consider first- and second- order Trotter formulas.
\begin{figure*}[t]
    \centering
    \includegraphics[width=0.9\linewidth]{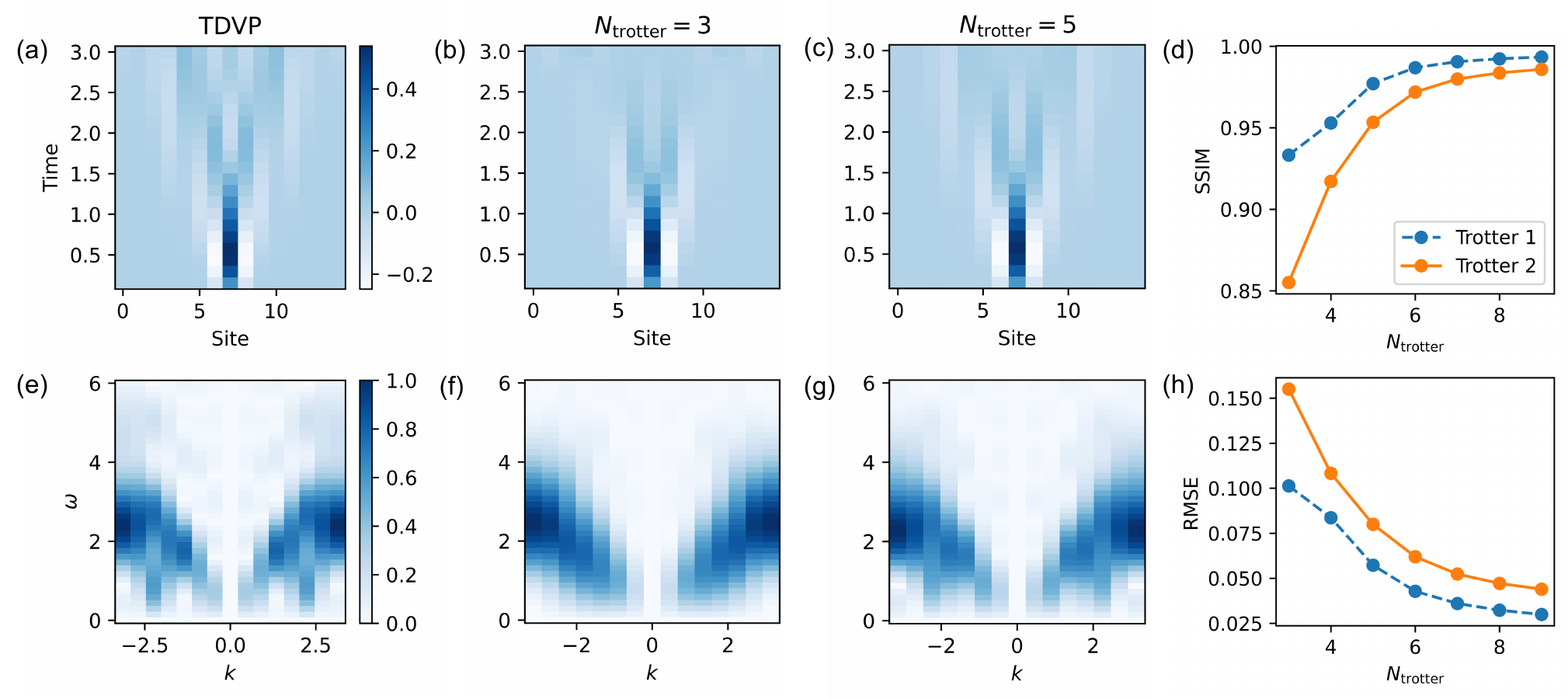}
    \caption{Trotterized time evolution of the 1D Fermi-Hubbard model with $L=15$, $U/J=3$ and $n_e=2/3$, using the ground state as the initial state. The latter has been obtained using the density-matrix renormalization group (DMRG) algorithm. The exact simulation has been performed using the TDVP algorithm, while the Trotterized simulations have been obtained with time-evolving block-decimation (TEBD) and expressing the Hamiltonian as a product of the interacting part and the hopping term on even and odd sites. DMRG and the time evolution algoritihms were executed with maximum bond dimension $\chi=64$. Top: (a) exact  and Trotterized time dynamics with (b) three and (c) five fixed Trotter steps using the first-order Trotter formula. Bottom: (e), (f), (g) absolute value of the QSF obtained by Fourier transforming and normalizing the time-dependent signals in panels (a), (b), (c), respectively. In the last column, (d) SSIM and (h) root mean square error (RMSE) between the QSFs obtained from exact and trotterized time-evolution simulations are shown for increasing numbers of Trotter steps. Both first- and second- order Trotter are simulated.
    }
    \label{fig:Trotter_simulation}
\end{figure*}
In view of implementation, it is crucial to determine the number of Trotter steps required to accurately resolve the spectrum. In some cases, a large number of steps may be necessary to reduce errors below a certain threshold, or the required number of steps might scale with the system size. Testing this is essential to ensure that the protocol remains viable for NISQ devices. In Fig. \ref{fig:Trotter_simulation}, we show that the Trotterization error in the spectrum reaches an acceptable threshold with just four Trotter steps for a reasonably large system size. Notably, the first-order Trotter method performs as well as (or even better than) the second-order method, allowing us to justify its use to minimise the depth of the Hamiltonian evolution ansatz.

The number of Trotter steps naturally depends on the total evolution time. For smaller systems, where long evolution times are unnecessary, the number of Trotter steps can be reduced while maintaining the same level of accuracy. In the following, the maximum number of Trotter steps that we use is $N_{\rm Trotter}=6$ (see Table~\ref{tab:exp_params}). 

In passing, we note that, alternatively, we could fix the Trotter step size, $\delta t$, instead of the number of Trotter steps, $N_{\text{Trotter}}$. Although fixing $\delta t$ may be beneficial to improve the quality of the space-time signal, maintaining a consistent level of noise across all time values is more effective when performing the Fourier transform. This consistency ensures that, after normalization, the Fourier transform removes any constant damping factors in a given observable (see Section~\ref{sec:hardware_results}). For this reason, we opted for the fixed Trotter steps approach.

\section{Quantum circuit implementation}
\label{sec:implem}
The encoding of fermions to qubits has an impact on the resources needed for the simulation. In this work, we use the Jordan-Wigner (JW) encoding, which maps every fermionic mode labeled by a site and spin index to one qubit. In this encoding, the creation and annihilation operators become~\cite{Cade_2020, Stanisic_2022}:
\begin{align}\label{eq:JW}
\begin{split}
    c_{i} &\rightarrow Z_0\dots Z_{i-1} \frac{X_i+iY_i}{2},\\
    c_{i}^\dagger &\rightarrow Z_0\dots Z_{i-1} \frac{X_i-iY_i}{2},
\end{split}
\end{align}
where the index $i = 0, ..., 2L-1$ now includes both the site and the spin degree of freedom. The hopping term between qubits $i$ and $j$ (with $i < j$) and on-site repulsion thus become
\begin{align}
\label{eq:JW_FH}
\begin{split}
    c_{i}^\dagger c_{j}+c_{j}^\dagger c_{i}& \rightarrow \frac{1}{2}(X_{i} X_{j}+Y_{i} Y_{j})Z_{i+1}\dots Z_{j-1},\\
    n_{i}n_{j}= c_{i}^\dagger c_{i}c_{j}^\dagger c_{j} &\rightarrow \frac{1}{4}(I-Z_i)(I-Z_j).
\end{split}
\end{align}
A common choice to minimize the weight of the Pauli operators corresponding to the hopping term of a fermionic Hamiltonian is to assign the first $L$ values to the modes with spin up and the last $L$ to the modes with spin down. In what follows, we will refer to this configuration as (all up)(all down). As we will show below, this may not be the optimal choice. 

\begin{figure*}
    \centering
    \includegraphics[width=\textwidth]{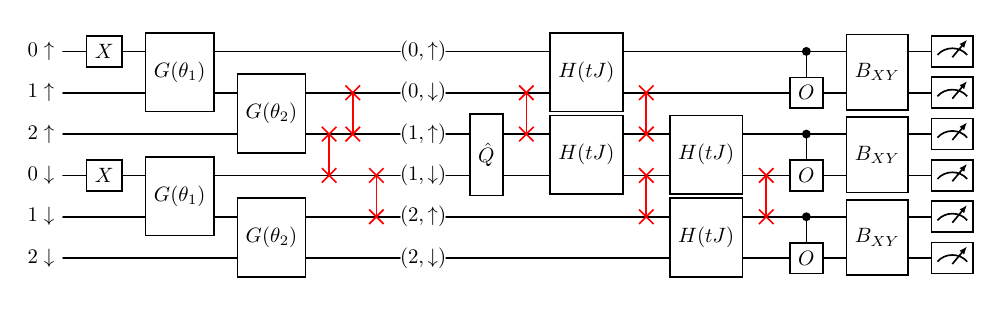}
    \caption{Complete circuit for a fermionic system of size $L=3$. Note the fermion-to-qubit ordering is changed after the initialisation with the Givens rotations. We first encode all the spin up modes in the first $L$ qubits and the spin down modes in the last $L$ qubits. We then move the qubits around with FSWAP gates (represented as swap gates in red) to encode every pair of modes $(i,\uparrow)$ and $(i, \downarrow)$ next to each other. After this reordering, the quench operator $\hat{Q}_j$ is applied to the two central qubits, corresponding to the spin up and down modes of fermion at $i=\lfloor L/2\rfloor$ site. The time-evolution operator is implemented using a first-order Trotter formula, consisting of even and odd hopping, and on-site interaction terms. Note that due to the (up, down)(up, down) JW ordering, implementing each hopping layers requires 3 layers of FSWAP gates per Trotter step. In the figure, only one Trotter step is shown. Finally, the unitary $B_{XY}$ diagonalizing the spin operator $\hat{S}^x_i$ is applied to each pair of qubits, which are then measured in the computational basis.}
    \label{fig:circuit}
\end{figure*}

As stated above, our simulation proceeds by the following steps. We start in the ground state of the free-fermion model $H_0$ (i.e., Eq.~\eqref{eq:FH} when $U=0$) for a given electronic occupation $N_e$ which can be prepared efficiently \cite{Verstraete09, jiang2018quantum}. As we discuss below, such a state can be prepared with a sequence of Givens rotations acting on pairs of adjacent modes of the same spin sector~\cite{jiang2018quantum}. A Givens rotation is a general rotation operation in the two-dimensional complex subspace. It is composed by a $\theta$-rotation to change the amplitude, and a $\phi$-phase operator. However, since we are dealing with real Slater determinants due to the time-reversal symmetry of the Hamiltonian, we can set $\phi = 0$. The rotation matrix for two adjacent modes labeled 1 and 2 is then given by
\begin{equation*}
    G(\theta) := 
    \begin{pmatrix}
        1 & 0 & 0 & 0 \\
        0 & \cos{\theta} & -\sin{\theta} & 0 \\
        0 & \sin{\theta} & \cos{\theta} & 0 \\
        0 & 0 & 0 & -1
    \end{pmatrix}=e^{i\frac{\theta}{2}(X_1Y_2-Y_1X_2)}.
\end{equation*}
The operation above can be implemented on two qubits by the rotation $G(\theta) = e^{-i\frac {\pi}{4}Z_2} N(\theta/2, 0, \theta/2)e^{i\frac {\pi}{4}Z_2}$, with $N(\alpha, \beta, \gamma) := e^{i(\alpha X_1 X_2 + \beta Y_1 Y_2 +\gamma Z_1 Z_2)}$. For $\beta = 0$, the two-qubit gate $N(\alpha, \beta, \gamma)$ can be implemented using two CNOTs as~\cite{Kraus_2001, Smith_2019}
\begin{equation*}
    \begin{quantikz} 
        \qw & \gate[2]{N(\alpha,0,\gamma)} & \qw\\
        \qw &  & \qw
        \end{quantikz}
        = 
        \begin{quantikz}
        \qw & \ctrl{1} & \gate{R_x(-2\alpha)} & \ctrl{1} & \qw\\
        \qw & \targ{} & \gate{R_z(-2\gamma)} & \targ{} & \qw
    \end{quantikz}.
\end{equation*}

Because the layers of Givens rotations are implemented to each spin sector independently, for this step it is convenient to map the spin up sector to the first $L$ qubits of the JW encoding and the spin down sector to the remaining qubits, i.e., choosing the (all up)(all down) ordering. 

Next, we need to find an implementation of the quench operator defined in  Eq.~\eqref{eq:quench_op_unitary} and applied to the central site.
The weight of this operator in the JW encoding can be very sensitive to the specific ordering we choose for the fermionic modes. Since we have chosen to start with an (all up)(all down) convention (which is convenient to minimize the depth of the Givens rotations circuit), the two qubits representing the spin up and down modes of the fermion at $j=\lfloor L/2\rfloor$ are not next to each other. In this case, the quench operator would have weight $ k = L + 1$ and would require a decomposition into $2\lceil\log_2(k)\rceil - 1$ two-qubit gates to be implemented~\cite{Clinton_2024}. Moreover, while the weight of the encoded on-site term of $H_{\rm FH}$ is not affected by the JW ordering (see Eq.~\eqref{eq:JW_FH}), it is important to keep in mind that in order to implement such interactions in a quantum computer with local connectivity, such as the IBM device we employ in this work, the qubits involved should be physically close to each other. In general, this is difficult to achieve using the (all up)(all down) JW ordering. To address both the issues, we can rearrange the qubits to the configuration (up, down)(up, down), where the spin up and spin down modes of each site $i$ are represented by consecutive qubits in the JW encoding. We also establish that this is the most resource-efficient ordering for our time-dynamics simulation (see Appendix~\ref{app:ordering} for other possible orderings). To switch from the (all up)(all down) to the (up, down)(up, down) ordering, we introduce a fermionic SWAP (FSWAP) network layer after the Givens rotations which rearranges the initial JW ordering such that the modes $(i,\uparrow)$ and $(i,\downarrow)$ are next to each other~\cite{Kivlichan_2018}. Using IBM's native gate set, each FSWAP can be decomposed as
\begin{equation*} 
    \begin{quantikz} 
        \qw & \gate[2]{\text{FSWAP}} & \qw\\
        \qw &  & \qw
    \end{quantikz}
    =
    \begin{quantikz}
        \qw &\gate{H}& \ctrl{1} & \targ{} & \qw & \qw \\
        \qw & \qw &\targ{} & \ctrl{-1}& \gate{H}& \qw
    \end{quantikz}.
\vspace{1em}
\end{equation*}
After applying the FSWAP network, the quench operator is local in the (up, down)(up, down) JW encoding and can be implemented in terms of qubit gates as
\begin{equation}
    \hat{Q}_j = e^{\frac{i\theta}{2}(X_{j} X_{j+1}+Y_{j} Y_{j+1})},
    \label{eq:quench_op_qubits}
\end{equation}
for $\theta = \pi/4$ and $j=\lfloor L/2\rfloor$. This unitary operator can be implemented on a quantum computer in the same way as the time-evolution operator of the hopping term of the Hamiltonian, $H_{12}(\theta):=e^{\frac{-i\theta}{2}(X_1X_2+Y_1Y_2)} = N(-\theta/2,0, -\theta/2)$ that, as discussed above, can be done using 2 CNOT gates. 

To implement the product formula of Eq.~\eqref{eq:Trotter}, we note that the even and odd hopping terms of the Hamiltonian $H_0$ in Eq.~\eqref{eq:FH}, $H_0^{\mathrm{even}}$ and $H_0^{\mathrm{odd}}$ (defined by constraining the summation over sites in Eq.~\eqref{eq:FH} to even and odd values of $i$, respectively), act on adjacent fermionic modes with the same spin, which, in our (up, down)(up, down) JW ordering, are represented by qubits that are one qubit apart. Consequently, we need an additional layer of FSWAPS both before and after each $H(\theta)$. On another hand, the time-evolution operator associated with the on-site interaction part of the Hamiltonian, $H_{\mathrm{int}}$, which act between two qubits representing the two spin modes of the same fermion site, can be implemented with the onsite gate $O(\phi)$,
\begin{equation*} 
    \begin{quantikz} 
        \qw & \ctrl{1} & \qw\\
        \qw &  \gate{O(\phi)} & \qw
    \end{quantikz}
    = 
    \begin{quantikz}
        \qw & \ctrl{1} & \qw\\
        \qw &\gate{e^{-i\phi}} & \qw
    \end{quantikz},
\end{equation*}
which in terms of CNOT gates becomes
\begin{equation*} 
    \begin{quantikz}
        \qw & \gate{R_z(\frac{\phi}{2})} & \ctrl{1} & \qw & \ctrl{1} & \qw & \qw \\
        \qw & \qw &\targ{} & \gate{R_z(-\frac{\phi}{2})} & \targ{} & \gate{R_z(\frac{\phi}{2})} & \qw
    \end{quantikz}.
\end{equation*}
Following the Trotter decomposition in Eq.~\eqref{eq:Trotter}, we repeat the application of the even and odd hopping terms, and the onsite term, $N_{\mathrm{Trotter}}$ times. 

After the Trotterized time evolution, we need to make the necessary measurements on the time-evolved state to infer the quantities of interest. The observable $\braket{\hat{S}^x_i} = \frac{1}{2}\braket{X_{i\uparrow}X_{i\downarrow}+Y_{i\uparrow}Y_{i\downarrow}}$ needs to be measured for each pair of fermionic modes $(i,\uparrow)$ and $(i,\downarrow)$. Here, $X_{i\sigma}$ and $Y_{i\sigma}$ denote Pauli operators acting on the qubit encoding the fermionic mode $(i,\sigma)$. Recalling that these modes are next to each other in the (up, down)(up, down) JW ordering of the qubits, we can apply the following unitary before the measurement~\cite{Cade_2020}
\begin{equation*} 
    \begin{quantikz} 
        \qw & \gate[2]{B_{XY}} & \qw\\
        \qw &  & \qw
    \end{quantikz}
    =
    \begin{quantikz} 
        \qw & \ctrl{1} & \gate{H} &\ctrl{1} & \qw\\
        \qw & \targ{}& \ctrl{-1}  & \targ{} & \qw
    \end{quantikz},
\end{equation*}
which can be implemented with 3 CNOT gates and diagonalizes $\hat{S}^x_i$ as $\ket{01}\bra{01} - \ket{10}\bra{10}$, allowing us to compute the expectation value of  $\hat{S}^x_i$ from the measurement results. 

Importantly, measuring the two-point correlators $\braket{X_{i}X_{i+1} + Y_{i}Y_{i+1}}$ using the above change of basis allows us to exploit the particle number conservation of the Fermi-Hubbard Hamiltonian for post-selection as a simple error mitigation technique~\cite{Stanisic_2022}. In particular, since in experiments on real quantum hardware we run many shots of the same circuit, we can retain only those bit strings containing the correct number of fermions, providing an effective way to filter out noisy shots.

In Fig. \ref{fig:circuit}, we show a simple circuit with all the components discussed in this section. 

\section{Approximate free-fermion initial state preparation}
\label{sec:approx_ff}
\begin{figure}
    \centering
    \includegraphics[width=0.7\linewidth]{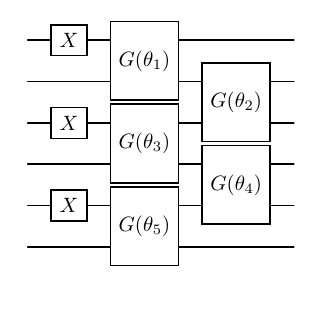}
    \caption{Approximate free-fermion initial state preparation for a Fermi-Hubbard model with $L=6$ in a single spin sector at half filling ($N_e = 3$) with a single layer of the DGA state preparation scheme. The gates $G(\theta_i)$ are Givens rotations whose angles are optimised via VQE.}
    \label{fig:schematic_approx_givens_layer}
\end{figure}

In the previous sections, we initialized the quench spectroscopy protocol to the free-fermion model ground state, as it provides a hardware-compatible alternative to the Fermi-Hubbard ground state for the circuit ansatz. However, the circuit to prepare the free-fermion ground state exactly with Givens rotations requires $O(L^2)$ gates at depth $O(L)$~\cite{jiang2018quantum}. In practical implementations, this depletes a significant amount of quantum resources. 
We propose here the Dense Givens Approximate state preparation scheme which is a simple, alternative approach that reduces quantum resource requirements while maintaining the quench spectroscopy algorithm's performance. Here, the free-fermion ground state is approximated with $n_{\mathrm{layers}}$ layers of Givens rotations. Each layer is maximally dense in Givens gates whereby every qubit is connected to its two nearest physical neighbors. Note that each layer has constant depth with respect to the system size. A schematic of such a single layer is shown in Fig.~\ref{fig:schematic_approx_givens_layer} for a single spin sector with $L=6$ at half-filling. The Givens angles are obtained by using the Variational Quantum Eigensolver (VQE) algorithm~\cite{Peruzzo_2014, Bharti_2022, Tilly_2022, Dalzell_2023} either by minimizing the energy with respect to the free-fermion Hamiltonian or by maximizing the overlap fidelity with the exact free-fermion ground state. For a desired occupation $N_e$, the fermions are initially maximally spread across both sectors, and this choice intuitively comes from the fact that the true ground state is homogeneous in particle density. 
For more analyses of DGA scheme's performance, see Appendix~\ref{app:DGA}. 

\subsection{Practical implementation}
\label{sec:DGA_practical}
\begin{figure*}
    \centering
    \includegraphics[width=0.9\linewidth]{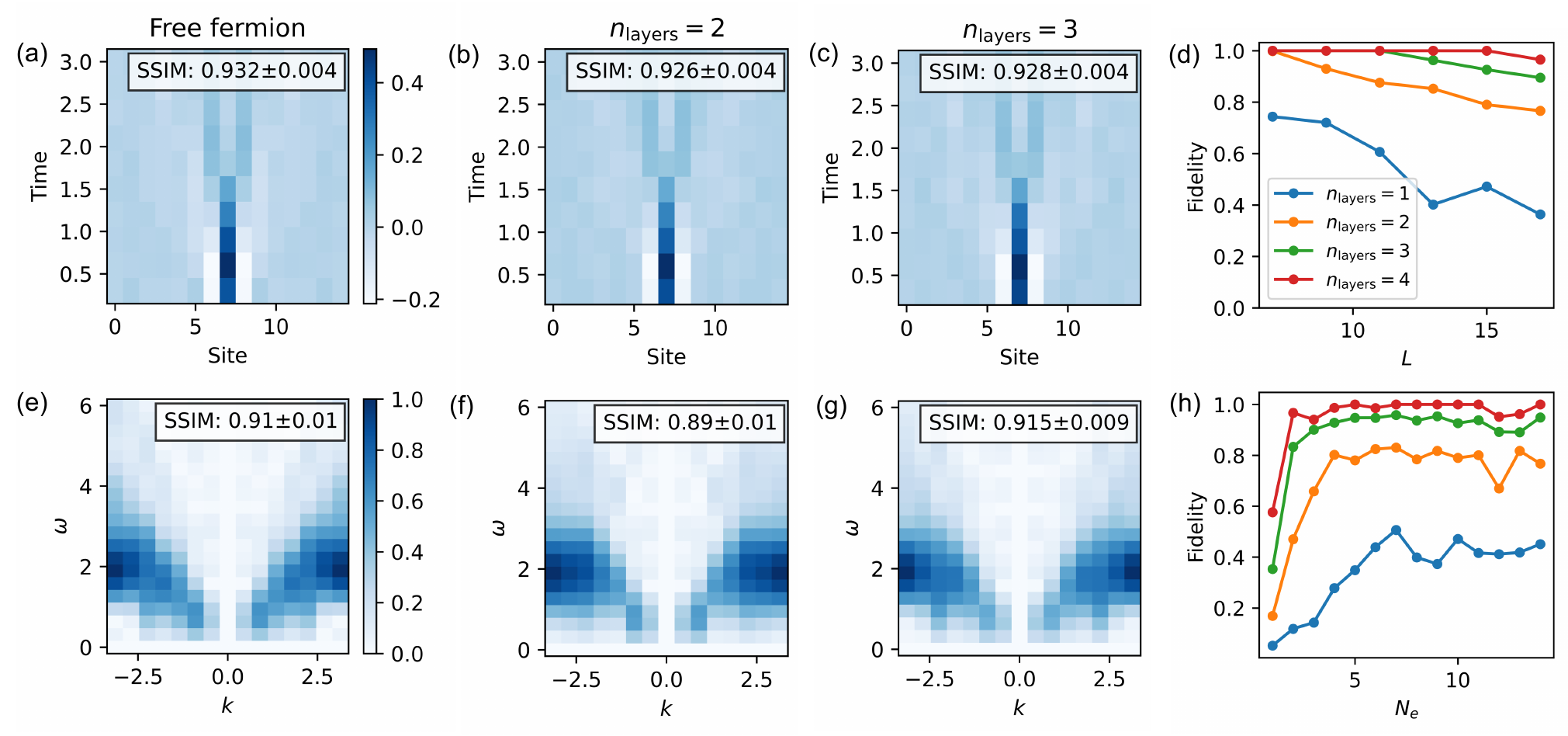}
    \caption{Trotterised time evolution with $N_{\text{Trotter}}=5$ Trotter steps of the Fermi-Hubbard model with $L=15$, $U=3$ and $n_e=2/3$, using the DGA approximation of the free-fermion ground state as the initial state. Top: time dynamics starting with the (a) free-fermion ground state and DGA free-fermion state with (b) two and (c) three layers. Bottom: QSFs corresponding to panels (a-c) obtained after a space-time Fourier transform. In the last column, the fidelity of the approximate DGA free-fermion state with respect to the exact free-fermion ground state is shown for different number of DGA layers, $n_{\rm layers}$. The fidelity is shown as a function of (d) the system size for a fermion filling fraction close to $n_e=2/3$ and (f) the electronic occupation $N_e$ for a fixed system size $L=15$.
    Data in panels (a-c) and (e-g) have been obtained by an exact simulation of the quantum circuits with $4096$ shots. Errors in the SSIM are computed using the bootstrap technique with $10^4$ instances~\cite{newman1999monte}.
    }
    \label{fig:approx_ff}
\end{figure*}
The advantage of using the DGA scheme is that we can calibrate the number of Givens rotation layers included in the approximation. Regardless of system size, each layer adds a fixed two-qubit depth of eight. Naturally, adding more layers improves fidelity, but our simulations show that for the system sizes considered in this work, no more than three layers are needed to achieve a SSIM $\geq 0.9$ when comparing the resulting spectrum to the exact one. This is shown in Fig.~\ref{fig:approx_ff}, where we compare the dynamics of $\braket{\hat{\boldsymbol{S}^x}(t)}$ and the QSF obtained starting the quench protocol from the free-fermion ground state (panels (a) and (e)), with the ones obtained using a DGA approximate state with $n_{\rm layers} = 2$ (panels (b) and (f)) and $n_{\rm layers} = 3$ (panels (c) and (d)). This allows us to significantly reduce the depth of the circuit without compromising the final outcome of the protocol, making the DGA state an efficient choice.

To evaluate the quality of the initial state approximation, we also examine the fidelity $F_{\rm DGA} = |\braket{\psi_{\rm DGA}|\psi_{\rm FF}}|^2$ between the approximate DGA state, $\ket{\psi_{\rm DGA}}$, and the free-fermion ground state $\ket{\psi_{\rm FF}}$, across different system sizes. As can be seen in Fig.~\ref{fig:approx_ff}(d), the fidelity for a given number of layers $n_{\text{layers}}$ decreases with increasing system size. 

We observe a clear correlation between the fidelity of the initial state and the SSIM of the resulting spectrum obtained at the end of the quench protocol: the better the initial state approximation, the closer the resulting spectrum is to the exact one (see Appendix~\ref{app:DGA}). 

As a result of the observations above, the number of layers required to maintain fidelity at an acceptable level increases with system size. However, this increase in circuit depth is significantly smaller compared to the growth that would result from using the full Givens rotation layers required for the exact free-fermion ground state preparation. For these reasons, we conclude that the DGA approximation of the free-fermion ground state serves as an effective initial state for our protocol, as it enables good (and, in some cases, better) results while requiring shallower circuits.

In Sec.~\ref{sec:classical_simulations} we have discussed how it is possible to retrieve a good approximation of the spin excitation spectrum of the 1D Fermi-Hubbard model using the unitary quench spectroscopy protocol by replacing the ground state of the Fermi-Hubbard model with the one of the corresponding free-fermion model. In this section, we have introduced the DGA scheme to prepare a high-fidelity approximation of the latter, allowing us to allocate the majority of the quantum resources to the time-evolution part of the protocol. It is important to note that, in principle, the state preparation part of the quantum circuit can be replaced by a VQE ansatz to directly prepare an approximation of the ground state of the Fermi-Hubbard model. This would make the initial state preparation a controlled approximation: as the number of VQE layers increases, the variational state is expected to converge to the real ground state~\cite{Bharti_2022, Tilly_2022, Cade_2020}. However, this approach would push the required resources beyond the capabilities of current hardware. For example, implementing the VQE algorithm to prepare the ground state of the Fermi-Hubbard model within the framework of the Hamiltonian Variational Ansatz (HVA) requires initializing the variational ansatz to the ground state of the corresponding free-fermion model~\cite{Wecker_2015b, Cade_2020, Stanisic_2022}. Therefore, in the next generation of NISQ devices, one can consider adding a few layers of the HVA after the DGA part of the circuit. We expect that that this will improve the efficacy of our algorithm in the strongly interacting regime.

\section{Results}
\label{sec:results}

\subsection{Simulations with different protocols}
\label{sec:results_sims}

\begin{figure*}
    \centering
    \includegraphics[width=0.9\linewidth]{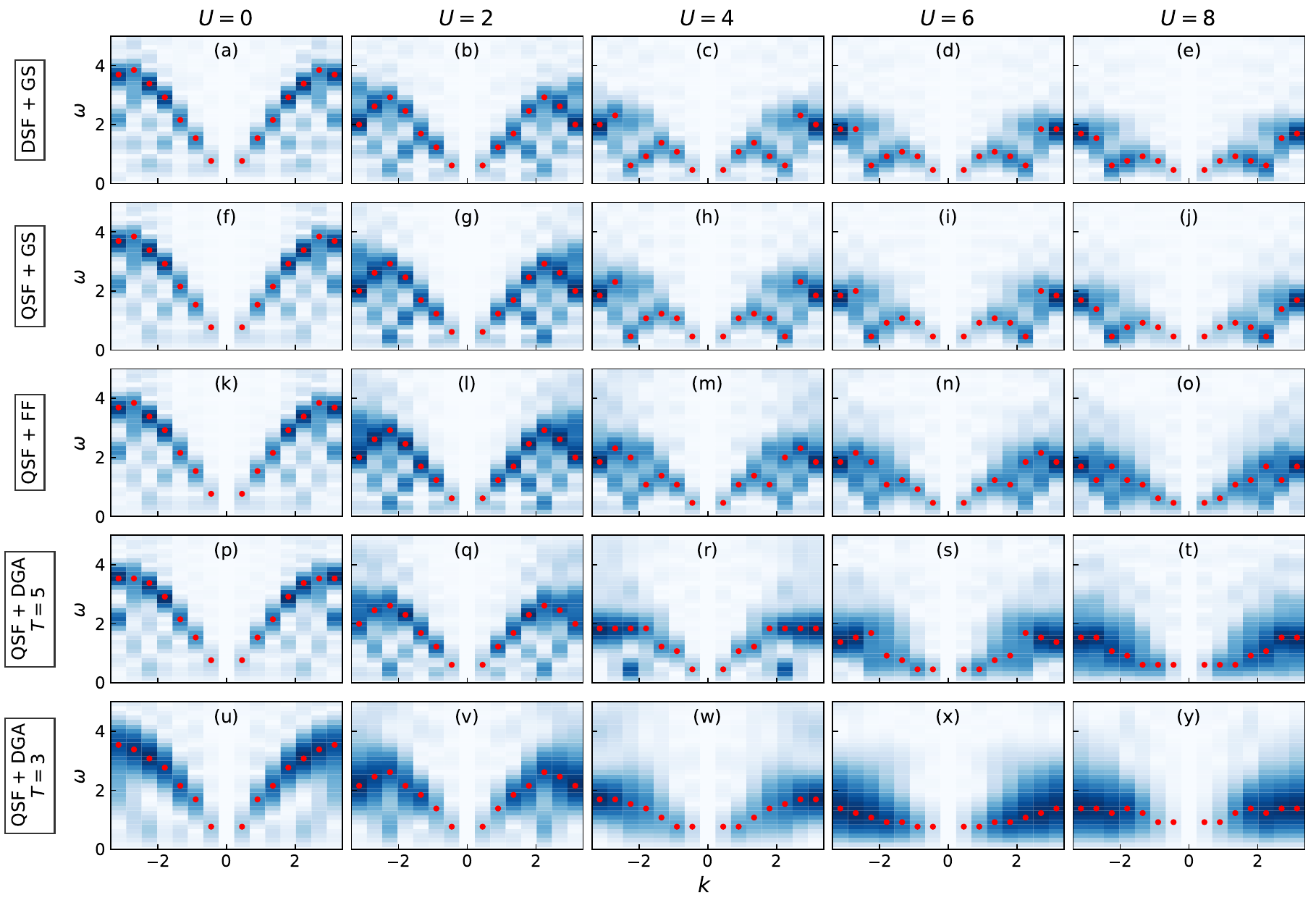}
    \caption{Dynamical (spin) structure factor (DSF) and QSFs for various protocols and values of $U$. Panels (a)-(e) show the exact DSF for the 1D Fermi-Hubbard model. Panels (f)-(j) display the QSF obtained by applying the unitary quench spectroscopy protocol to the ground state (GS) of the 1D Fermi-Hubbard model, (k)-(o) show results for a quench from the free-fermion (FF) model ground state, and (p)-(t) correspond to a quench from the DGA state with $n_{\rm layers} = 3$ using a Trotterized dynamics with $N_{\rm Trotter} = 10$ Trotter steps. All results in (u)-(y) are obtained by taking the absolute value of the normalized Fourier transform of $\braket{\hat{\boldsymbol{S}^x}(t)}$ after a total evolution time $T=5$ on a system with $L = 15$. The final row (p)-(t) shows the QSF obtained using the DGA approximate state, using $T=3$, $n_{\rm layers} = 3$ DGA layers, and $N_{\rm Trotter} = 6$ Trotter steps, matching the settings used for hardware results discussed in Section~\ref{sec:hardware_results}. In all panels, the red dots denote the maximum of the QSF for each value of $k$. Data in panels (a)-(j) have been obtained using the TDVP algorithm with maximum bond dimension $\chi=128$. Data in panels (k)-(t) are obtained by simulation of the quantum circuit with $4096$ shots.}
    \label{fig:comparative_panel}
\end{figure*}

In this section, we begin by presenting a comprehensive analysis of the two-spinon spectrum for the 1D Fermi-Hubbard model obtained with different approaches and on-site interaction strengths. The results are shown in Fig.~\ref{fig:comparative_panel}. 
Here, the first row displays the exact dynamical spin structure factor of the 1D Fermi-Hubbard model, while the second row shows the QSF obtained starting the quench spectroscopy protocol from the ground state of the Fermi-Hubbard model, $\ket{\Psi}$, as given by Eq.~\eqref{eq:GF_quench} with $\theta = \pi/4$. As anticipated in Sec.~\ref{sec:protocol}, the agreement between the two quantities is excellent. As discussed in Section~\ref{sec:classical_simulations}, one possibility to reduce the resources to implement the protocol on a real quantum device is to apply the quench to the ground state of the free-fermion model, $\ket{\psi_{\rm FF}}$, in order to save computational resources for the time-dynamics simulation. As discussed there, we expect this protocol to give a good approximation of the system's excitation spectrum if only a single spin excitation (i.e., a single pair of spinons) is present in the quenched state, which occurs for small values of $U$ (see Fig.~\ref{fig:free_fermion_simulation}(a)). By comparing the second and third rows in Fig.~\ref{fig:comparative_panel}, we can see that the QSF computed using the Fermi-Hubbard ground state and the one obtained by applying the quench to the free-fermion ground state are in excellent agreement for $U < 4$. Remarkably, the agreement between the two QSFs remains good even for larger value of $U$, where all the main features of the dynamical spin structure factor are still visible in the approximate QSF (such as the softening of the spin excitations for large $U$ and the cusp with $\omega\rightarrow0$ for $k\rightarrow \pm 2k_F$). In particular, we notice that the main discrepancies arise for $k\rightarrow \pm 2k_F$. This is expected since for these values of $k$ spin excitations become gapless and many pairs of spinons with $k\approx\pm 2k_F$ can potentially be excited with a vanishing amount of energy, leading to a breakdown of the quench spectroscopy protocol. 

As discussed above, we can use the DGA state as an alternative to the free-fermion ground state to further reduce circuit depth and number of two-qubit gates, making it more suitable for near-term hardware implementations. As long as the fidelity of the DGA state, $\ket{\psi_{\rm DGA}}$, with respect to the free-fermion ground state is sufficiently high, the QSF computed using the DGA state captures all the key spectral features of the spin excitation spectrum with an accuracy comparable with the QSF obtained using the free-fermion ground state for $U<6$. Note that in panels (p)-(y) of Fig.~\ref{fig:comparative_panel} we also incorporate the effects of Trotterization, which are the major responsible for the loss of quality of the approximate QSF for larger values of $U$.

Finally, the last row of Fig.~\ref{fig:comparative_panel} shows the QSF computed using the DGA initial state and the parameters that we used in our experiments on IBM hardware, described in Section~\ref{sec:hardware_results} below.
Here, in order to reduce circuit depth and two-qubit gate counts, we run simulations up to $T=3$ using six Trotter steps. These parameters are sufficient to recover a satisfactory spin excitation spectrum for $U=3$ (see Fig.~\ref{fig:real_hardware}). On the other hand, to access larger values of $U$ more Trotter steps and a longer time evolution are needed. These observations emphasize the balance required between initial DGA state fidelity, circuit depth and two-qubit gate counts, and time-evolution accuracy to optimize the performance of the unitary quench protocol.

\subsection{Resource estimation}
Following the gate decomposition and circuit ansatz described in Section~\ref{sec:implem}, we can determine the gate count and circuit depth. These metrics depend on the chosen qubit ordering. For example, operations like the quench must be applied to adjacent qubits to avoid handling the JW string. Therefore, if the selected ordering places modes $(\lfloor L/2\rfloor, \uparrow)$ and $(\lfloor L/2\rfloor, \downarrow)$ on distant qubits, as in the (all up)(all down) encoding we discussed in Section~\ref{sec:implem}, we need to introduce a layer of FSWAPs to bring them together, which increases both gate count and circuit depth. Specifically, each FSWAP layer adds a factor of $O(L)$ to the depth. 

As discussed, due to the constraints and characteristics of the IBM hardware, we prioritize the (up, down)(up, down) ordering shown in Fig.~\ref{fig:circuit} to perform the quench, the time evolution, and the measurement part of the unitary quench spectroscopy protocol. However, we start from the (all up)(all down) ordering in which the initialization part of the circuit becomes straightforward, as the Givens rotations are applied in independent layers to the spin-up and spin-down sectors. Specifically, for the initialization to the free-fermion ground state, $\ket{\psi_{\rm FF}}$, we have $N_G = (L - N_{e}/2) N_{e}$ gates \cite{jiang2018quantum}. This number is maximized at $L^2/2$ in the half-filled case, where $N_{e} = L$. 
We use this maximum to establish an upper bound on the gate count for any initial state. This gate block consists of $L - 1$ layers of Givens gates, and since each Givens rotation can be implemented with a depth of $2$ two-qubit gates, this section of the circuit has a total depth of $2(L - 1)$. As discussed above, we can instead use an $n_{\text{layers}}$ DGA approximation of the free-fermion ground state that brings the depth of this part of the circuit down to $4n_{\text{layers}}$. We connect every pair of fermions in every Givens layer (as in Fig.~\ref{fig:schematic_approx_givens_layer}) so the total number of two-qubit gates in this case is $2(L-1)n_{\text{layers}}$.

We then require an FSWAP layer consisting of $(L^2 - L)/2$ gates and a two-qubit gate depth of $2(L - 1)$ to rearrange the qubits into the (up, down)(up, down) ordering. The subsequent quench operator adds only a two-qubit gate depth factor of two. Likewise, each Hamiltonian evolution gate has a depth of two. Each Trotter step consists of three layers (even and odd hopping, and on-site interaction terms) that can be implemented simultaneously. Taking into account the FSWAPs applied before and after the each hopping term, the total two-qubit gate depth for the time evolution part amounts to $12N_{\text{Trotter}}$.

Finally, the unitary operator $B_{XY}$, which diagonalises the spin operator $\hat{S}_i^x$, is applied to each pair of qubits. As $B_{XY}$ can be implemented with three CNOT gates, this final layer contributes a constant two-qubit gate depth of $3$ and a total of $3L$ two-qubit gates.

The advantage of the (up, down)(up, down) encoding is that it eliminates the need to rearrange the qubits before measurement, as all modes associated with the same fermionic sites are already paired adjacently. Additionally, in this scheme, the on-site interaction gates are applied locally, meaning no additional CNOT layers are required when compiling to a real device with local connectivity. Consequently, the total circuit depth sums to $2L+12N_{\text{Trotter}}+4n_{\text{layers}}+3$, and the number of two-qubit gates is $L^2 + L(10N_{\text{Trotter}}+2n_{\text{layers}}+2) + (2 - 8N_{\text{Trotter}}-2n_{\text{layers}})$. 

\subsection{Quantum hardware results}
\label{sec:hardware_results}
In this section, we present the spin excitation spectrum of the 1D Fermi-Hubbard model obtained using the unitary quench spectroscopy algorithm on IBM quantum hardware. The circuit in Fig.~\ref{fig:circuit} is implemented starting from the approximate free-fermion state obtained via the DGA scheme described in Section~\ref{sec:approx_ff}, with $n_{\text{layers}}=2$ for $L \in \{9,11,13\}$, and $n_{\text{layers}}=3$ for $L=15$. The number of layers for the DGA scheme is chosen to ensure that the fidelity between the DGA and the free-fermion ground state is close to a threshold of 0.9 (see Table~\ref{tab:exp_params} for details). Interestingly, for certain system sizes, we find that even when the fidelity falls below this threshold, the spectrum is still recoverable, yielding excellent results. Although some spectral features may appear weaker when using a shallower DGA ansatz with lower fidelity, they can often be enhanced by increasing the number of Trotter steps as discussed in Section~\ref{sec:results_sims}. After initialization, we apply the FSWAP layer, the quench operator, the Trotterised time evolution, and finally we measure in the $XX+YY$ basis, enabling particle number post-processing (see Section~\ref{sec:implem}). In order to mitigate noise in the quantum simulations, we employ techniques such as an $XX$ dynamical decoupling sequence~\cite{Viola98,Pokharel18}, and Pauli twirling on two-qubit gates and measurements~\cite{Wallman16} as implemented in the Qiskit SDK~\cite{JavadiAbhari_2024}. In particular, we expect the synergy between the quench spectroscopy protocol and Pauli twirling to make the algorithm very robust to noise. Pauli twirling converts coherent noise into Pauli noise~\cite{Wallman16, Hashim_2021} which effectively acts on single-site Pauli operators $P_i$ as a depolarizing channel, $\braket{P_i(t)}_{\rm noisy} = [1-p_i(t)]\braket{P_i(t)}$, with $p_i(t)$ a time-dependent error rate. If we assume that $p_i(t) \approx p$, $\forall i, t$, then the noisy space-time signal retrieved from a quench experiment will be $\braket{\hat{\boldsymbol{S}^x}(t)}_{\rm noisy} \approx (1-p)\braket{\hat{\boldsymbol{S}^x}(t)}$, since $\braket{\hat{\boldsymbol{S}^x}(t)}$ consists of single-site Pauli operators. Since the QSF is normalized so that the maximum of its absolute value is one, the constant damping factor $(1-p)$ does not affect its features, regardless of the damping of the space-time signal. To ensure that $p_i(t)$ is as constant as possible for the different values of $t$, in this work we use the same number of Trotter steps $\forall t\in[0,T]$. On the other hand, thanks to Pauli twirling, we expect $p_i(t)$ to be homogenous on all qubits to a good degree of approximation.

\begin{figure*}
    \centering
    \includegraphics[width=0.9\linewidth]{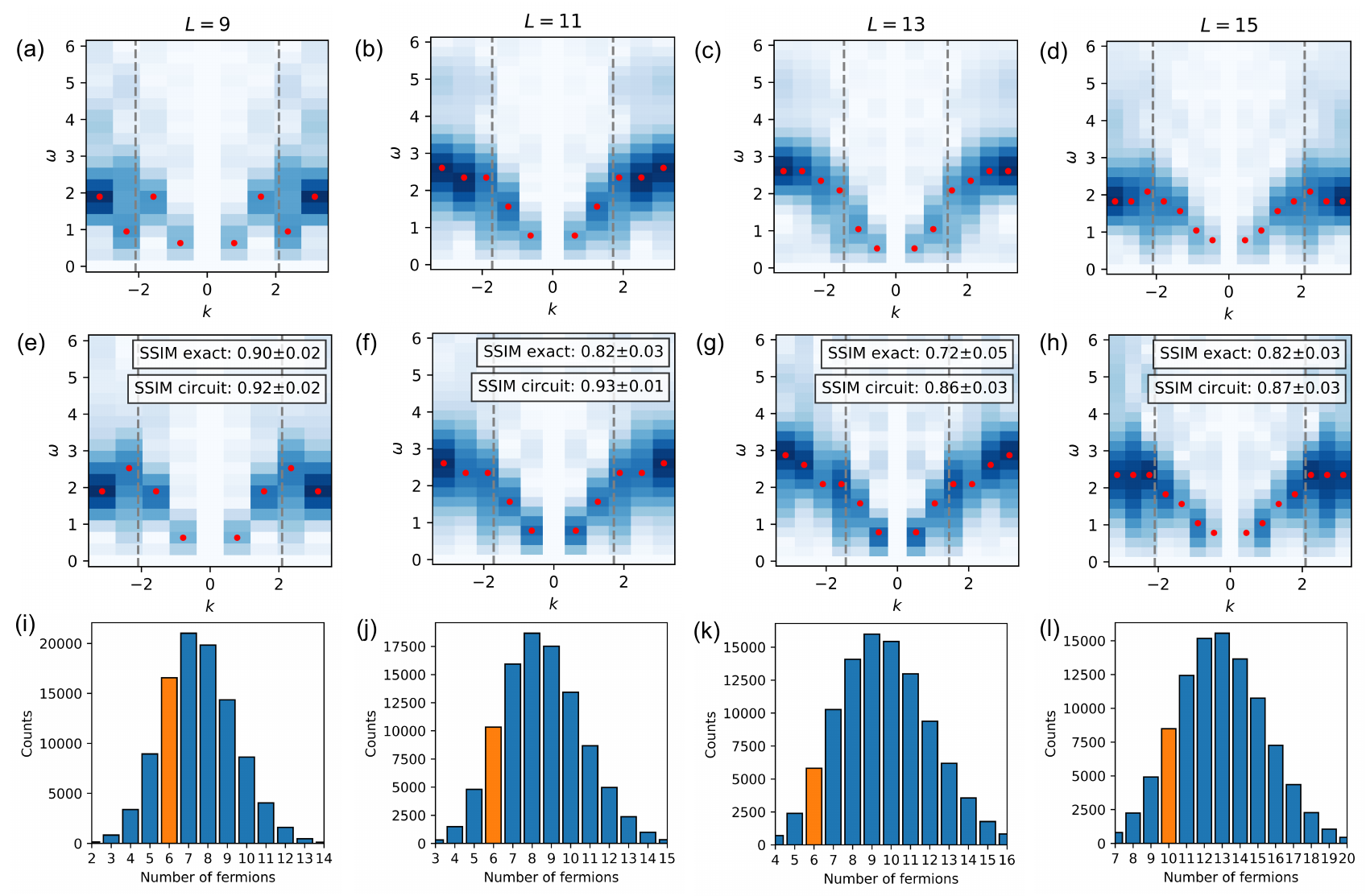}
    \caption{Simulation and hardware results from quench spectroscopy simulations of the 1D Fermi-Hubbard model for system sizes $L=9, 11, 13, 15$ (from left to right). The circuit details and the parameters used in these experiments are reported in Tables~\ref{tab:resources} and \ref{tab:exp_params}, respectively.
    (a)-(d) Spectra from noiseless simulations of the quantum circuits. (e)-(h) Spectra from experiments. All the circuits are executed on the IBMQ device ibm\_fez using a total of $10^6$ shots and $8$ Pauli twirling randomizations per circuit.
    (i)-(l) Histograms showing the distribution of counts across different fermion numbers at the end of the computation for the same system sizes, with the initial fermion count $N_e$ highlighted in orange. In panels (a-h), the theoretical position of the cusp of the spin excitation spectrum at $2k_F = \pm \pi n_e $ is marked by a dashed line, while the red dots denote the maximum of the QSF for each value of $k$. In panels (e-h) errors on the SSIM have been computer using the bootstrap technique over $10^4$ instances~\cite{newman1999monte}. 
    }
    \label{fig:real_hardware}
\end{figure*}

Figure~\ref{fig:real_hardware} shows the results obtained from implementing the unitary quench spectroscopy circuit on the IBM quantum processor ibm\_fez. 

The recovered spectra are in excellent agreement with the theoretical predictions for all the system sizes we considered, ranging from $L=9$ (with $18$ qubits) to $L=15$ (with $30$ qubits).
As the system size $L$ increases, errors in the computation due to hardware noise become more apparent. As observed in the particle number histograms in panels (i-l), computational errors can cause the inclusion of spectral signals from unintended particle number sectors, potentially interfering with the reconstruction of the spin excitation spectrum. The distribution of counts in the histograms is compatible with noise driving the system to a maximally mixed state, where the bit strings distribute as a Gaussian around half filling $N_e = L$. On the other hand, readout errors on short circuits, where the system state does not become fully mixed, could explain the skewness of the distribution. In this work we exploit the particle number conservation of the Fermi-Hubbard model to post-select only those bit strings containing the correct number of fermions (see the orange bars in the histograms in Fig.~\ref{fig:real_hardware}(i-l)). This results in an overall improvement of the quality of the reconstructed spectra. However, we also note that, as we discuss in Appendix~\ref{app:depolarizing}, the protocol has some intrinsic resilience to bit-flip errors, which are the main responsible for the change in fermionic particle numbers. This fact makes this approach suitable even for trapped ions and neutral atoms architectures, where collecting a large number of samples is often impractical due to their inherently slower operation speed~\cite{Hemery_2023}.

Due to the nature of the noise and the small depth of the circuit, we expect errors in the signal $\braket{\hat{\boldsymbol{S}^x}(t)}$ to be localized in space and time. In this case, they would mainly affect the short-wavelength/high-energy region of the spectrum, as can be seen, for instance, from the vertical stripes at large momenta $k \approx \pm\pi$ on the sides of the $L=13$ and $L=15$ spectra in panels (g) and (h) of Fig.~\ref{fig:real_hardware}, respectively. Despite this, we can successfully recover the main features of the spin excitation spectrum, such as correctly identifying the cusp for $k=\pm 2k_F$ and locating the maximum of the (absolute value of the) QSF as a function of $k$. In addition to the long-wavelength/high-energy noise, we also note some discrepancies in the position of the maximum of the QSF in the recovered spectra for $k\approx\pm 2k_F$. In particular, see the case with $L=9$ in panels (a) and (e) in Fig.~\ref{fig:real_hardware}. These can be explained by noting that even in the simulated results in panels (a-d) the QSF for $k\approx\pm 2k_F$ is spread over a large window of energy values. Therefore, we expect the position of the maximum of the QSF to be sensitive to small fluctuations in the space-time signal. Moreover, as discussed at the beginning of Section~\ref{sec:results}, in this region the spin excitations are almost gapless and deviations from the two-spinon continuum are more likely due to the presence of a potentially larger number of spin excitations.

\begin{table}[t!]
    \centering
    \begin{tabular}{ c c c c } 
    \hline \hline
     $L$ & $N$ & Depth & Gates \\
     \midrule
     9  & 18 & 89 & 543 \\ 
     11 & 22 & 105 & 797 \\ 
     13 & 26 & 109 & 977\\ 
     15 & 30 & 117 & 1193\\ 
     \hline \hline
    \end{tabular}
    \caption{Number of qubits, two-qubit depth and total number of two-qubit gates for the experiments shown in Fig.~\ref{fig:real_hardware}.
    }
    \label{tab:resources}
\end{table}
    
\begin{table}[t!]
    \centering
    \begin{tabular}{ c c c c c c} 
    \hline\hline
     $L$ & T & $N_e$ & $n_{\mathrm{layers}}$ & $F_{\rm DGA}$ & $N_{\mathrm{Trotter}}$ \\
     \midrule
     9  & 3 & 6 & 2 & 0.93 & 5 \\ 
     11  & 3 & 6 & 2 & 0.90 & 6\\ 
     13  & 3.5 & 6 & 2 & 0.88 & 6\\
     15  & 3 & 10 & 3 & 0.95 & 6 \\
     \hline\hline
    \end{tabular}
    \caption{System size $L$, maximum evolution time $T$, fermion occupation number $N_e$, number of Trotter steps $N_{\mathrm{Trotter}}$, number of DGA layers $n_{\mathrm{layers}}$, and fidelity between the approximate DGA state and the free-fermion ground state for the experiments shown in Fig.~\ref{fig:real_hardware}.}
    \label{tab:exp_params}
\end{table}

Note that for the simulation with $L = 13$ we choose a smaller occupation number $N_e$ because due to the limited momentum resolution even the spectrum obtained from a noiseless simulation does not match the one obtained from the corresponding TDVP simulation. We expect this issue to become negligible for large system sizes. On the other hand, we take advantage of this change in occupation number to show how the protocol is capable of correctly capturing the softening of the spin excitations at lower densities even when it is executed on quantum hardware. 

\section{Discussion}

In this work we introduced the unitary quench spectroscopy protocol which allowed us to obtain the spectral properties of the spin excitations of a 1D Fermi-Hubbard model from a digital quantum processor, without resorting to auxiliary qubits or extra measurements. We exploited the intrinsic robustness of the protocol to perturbations to the initial state to significantly reduce the resources required to prepare the initial state. This has been achieved by introducing the DGA scheme to prepare a high-fidelity approximation of a free-fermion ground state using a shallow circuit with depth independent of the system's size. Crucially, this allowed us to save the majority of the resources for the time-dynamics simulations. 

Since spectral properties of quasiparticle excitations are encoded into the peaks of the corresponding retarded Green's function, dynamical structure factor, or quench spectral function, they can be extracted from normalized signals. This makes it possible to effectively mitigate the effects of global depolarizing noise. In this context, we demonstrated that the unitary quench spectroscopy protocol is extremely effective when combined with simple noise-tailoring techniques, such as Pauli twirling, which turns an unknown noise channel into a Pauli channel. All of this, combined with a careful choice of the Jordan-Wigner ordering of the fermionic modes, allowed us to extract useful information about the spin excitation spectra of large instances of the 1D Fermi-Hubbard model~\cite{Arute_2020, Stanisic_2022, Bespalova_2024, Hemery_2023}. Moreover, the two-qubit gate depth and count of the circuits we considered are not far from state-of-the-art experiments implementing sophisticated error mitigation techniques~\cite{Kim_2023, RobledoMoreno_2024}.
We have therefore established that even in the presence of noise, useful spectroscopic information of fermionic systems can be extracted from noisy quantum devices.

\begin{acknowledgments}
We would like to thank Jan Lukas Bosse, Charles Derby, Joel Klassen, Harry McMullan, and the rest of the Phasecraft team for helpful discussions. We also thank the IBM Quantum Support team for the assistance and Aleksei Ivanov for pointing out an issue in an older version of the manuscript. The project has received funding from the InnovateUK grant 10032332 and the InnovateUK grant 44167. Data supporting the figures in this manuscript are available at Ref.~\cite{data}.
\end{acknowledgments}


\appendix

\section{Exact retarded charge Green's function from a double quench experiment}
\label{app:charge_response}
\begin{figure}
    \centering
    \includegraphics[width=0.9\columnwidth]{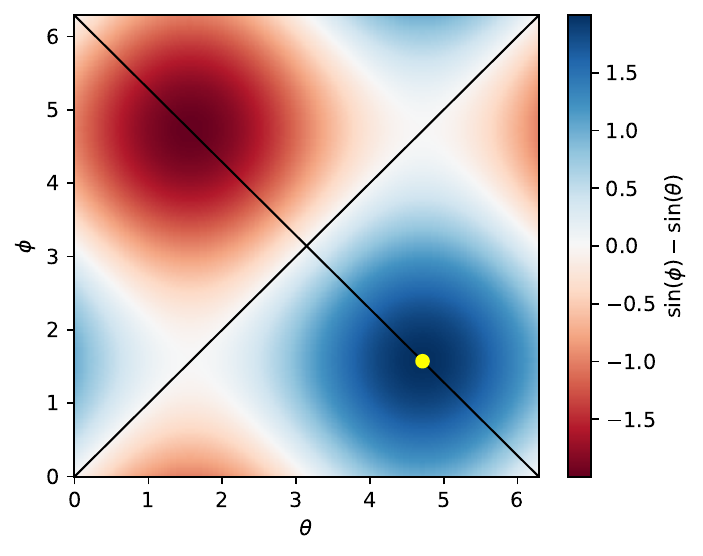}
    \caption{Density plot of the coefficient of $G^n_{jk}(t)$ in the double quench spectroscopy protocol of Eq.~\eqref{eq:GF_double_quench_charge}, $\sin\phi-\sin\theta$. The black contours denote the region of the $\theta-\phi$ plane in which Eq.~\eqref{eq:cos_condition_charge} is satisfied. The yellow dot, corresponding to $(\theta, \phi) \approx (3\pi/2, \pi/2)$, represent an optimal pair of angles for the protocol.}
    \label{fig:double_quench_conditions_charge}
\end{figure}

Here, we show that the double quench protocol discussed in Section~\ref{sec:protocol} allows one to access the exact retarded charge Green's function of a fermionic system (and, therefore, its dynamical charge structure factor)~\cite{Schuckert_2020_probing}, 
\begin{equation}
    \label{eq:charge_GF}
    G^n_{jk}(t) := -i\bra{\Psi}[n_k(t),n_j]\ket{\Psi},\quad \text{for } t>0.
\end{equation}
Let us consider the following quench operator,
\begin{equation}
    e^{i\theta n_j} = 1+(e^{i\theta}-1)n_j,
\end{equation}
where we have used that $(n_j)^k = n_j$. The dynamics of $n_k(t)$ on the quenched state
\begin{equation}
    \ket{\overline{\Psi}^j_\theta} = e^{i\theta n_j} \ket{\Psi}= \left[1+(e^{i\theta}-1)n_j\right]\ket{\Psi}.
\end{equation}
is given by
\begin{align}
     \bra{\overline{\Psi}^j_\theta} n_k(t)\ket{\overline{\Psi}^j_\theta} &= \bra{\Psi} n_k(t)\ket{\Psi} \nonumber\\
     &+(\cos\theta-1)\bra{\Psi}\{n_k(t),n_j\}\ket{\Psi} \nonumber\\
     &+ i\sin\theta\bra{\Psi}[n_k(t),n_j]\ket{\Psi} \nonumber\\
     &+ 2(1-\cos\theta)\bra{\Psi}n_jn_k(t)n_j\ket{\Psi},
\end{align}
where $\{A,B\}: = AB + BA$ is the anticommutator. In the third row, one can recognize the retarded charge Green's function of Eq.~\eqref{eq:charge_GF}. To extract it, we consider two quenches with angles $\theta$ and $\phi$ such that
\begin{subequations}
    \label{eq:double_quench_conditions_charge}
    \begin{align}
        \sin\theta -\sin\phi &\neq 0,\\
        \cos\theta-\cos\phi & = 0. \label{eq:cos_condition_charge}
    \end{align}
\end{subequations}
As can be seen from Fig.~\ref{fig:double_quench_conditions_charge}, there exist infinitely many pairs $(\theta,\phi)$ satisfying Eqs.~\eqref{eq:double_quench_conditions_charge}. By picking one of them, the exact retarded charge Green's function can then be obtained as
\begin{align}
    \label{eq:GF_double_quench_charge}
    &\bra{\overline{\Psi}^j_\theta} n_k(t)\ket{\overline{\Psi}^j_\theta} - \bra{\overline{\Psi}^j_\phi} n_k(t)\ket{\overline{\Psi}^j_\phi} \nonumber \\
    &= (\sin\phi-\sin\theta)G^n_{jk}(t).
\end{align}
The yellow dot in Fig.~\ref{fig:double_quench_conditions_charge}, corresponding to $(\theta, \phi) = (3\pi/2, \pi/2)$, shows an optimal choice for the pair $(\theta,\phi)$, which maximizes the value of the coefficient of $G^n_{jk(t)}$ in Eq.~\eqref{eq:GF_double_quench_charge}, i.e., $\sin\phi - \sin\theta = 2 $.

\section{Structural similarity index measure}
\label{app:ssim}
In this work, in order to make quantitive comparisons between spin excitation spectra, we use the structural similarity index measure (SSIM), which is a well-established metric to quantify the similarity between two images in the field of image processing~\cite{Zhou_2004}. The SSIM between to images $x$ and $y$ is defined as
\begin{equation}
\label{eq:app:ssim}
\mathrm{SSIM}(x, y) := \frac{(2\mu_x \mu_y + C_1)(2\sigma_{xy} + C_2)}{(\mu_x^2 + \mu_y^2 + C_1)(\sigma_x^2 + \sigma_y^2 + C_2)},
\end{equation}
where $\mu_\nu:=N^{-1}\sum_i\nu_i$ is the mean intensity of $\nu = \{x,y\}$, $\sigma_\nu^2$ is its variance, and $\sigma_{xy}$ is the covariance between $x$ and $y$. The SSIM can also be written as a product of luminance similarity, contrast similarity, and structural similarity, providing a score between $-1$ and $1$, where $1$ indicates perfect similarity between the images. In Eq.~\eqref{eq:app:ssim}, the constants $C_1$ and $C_2$ are small positive values used to stabilize the division when the denominator is close to zero. In practice, they are chosen as $C_1 = (K_1R)^2$ and $C_2 = (K_2R)^2$, with $K_1 = 0.01$ and $K_2=0.03$~\cite{Zhou_2004}. Here, $R$ is the dynamic range of the images, i.e., the difference between the maximum and minimum possible values in the image domain. Since QSFs have values in $[0,1]$, in this work we have fixed $R = 1$. We have used the same value of $R$ for plots of $\langle \hat{S}^x_i(t)\rangle$, as in all the cases we considered the dynamic range of this observable is approximately 1.

\section{Zero-padding and Fourier transforms}
\label{app:fourier}
In this work, all the Fourier transforms are computed using the zero-padding technique~\cite{Lyons_2011}. For a given time signal $f(t)$, with $t\in[0,T], $ $N_\mathrm{samples}$ are taken with sampling rate $f_s := N_\mathrm{samples}/T$. The corresponding spectral bandwidth is $\omega\in[-\pi f_s, \pi f_s]$ and the spectral resolution is $\Delta_\omega := 2\pi/T$. To better resolve the features of the spectra, we use the zero-padding technique and append $N'_\mathrm{samples} - N_\mathrm{samples}$ zeros to the original signal, artificially extending the simulation time from $T$ to $T':=(N'_{\rm samples}/N_{\rm samples})T$. The new signal, $f'(t)$ satisfies $f'(t) = f(t)$ for $t\in[0,T]$ and $f'(t)=0$ for $t\in(T,T']$. The spectral resolution of the padded signal is then given by $\Delta'_\omega = 2\pi/T'$. Of course, this only improves the quality of the original signal by smearing the spectral bins of the original signal and does not allow one to resolve features with spectral width smaller than the original resolution $\Delta_{\omega}$. Finally, since the system's Hamiltonian is time-translation invariant, we can improve the spectral resolution by a factor of 2 by reflecting the signal around $t=0$. We obtain a signal $f''(t)$ for $t\in[-T',T']$ such that $f''(-t) = f''(t)$ and $f''(t) = f'(t)$ for $t>0$. The spectral resolution of the new signal is therefore $\Delta''_\omega = \pi/T'$.

In this work, all features of the spin excitation spectra for the parameters we consider are contained in the spectral interval $\omega\in[-\omega_\mathrm{max},\omega_\mathrm{max}]$, with $\omega_\mathrm{max} = 6J$. Heuristically, we obtain good results by choosing the value of $N'_\mathrm{samples}$ so that the interval $[0,\omega_{\mathrm{max}}]$ is divided into $2N_\mathrm{samples}$ frequency bins. This implies $\omega_\mathrm{max} = 2\pi N_\mathrm{samples}/T'$ and therefore $N'_{\rm samples} = 2\pi N_{\rm samples}^2/(\omega_{\rm max}T)$.

\section{(all up)(all down) JW ordering}
\label{app:ordering}

In the main text, we employ the (up, down)(up, down) JW ordering to encode fermionic modes into qubits. However, alternative encodings may be more efficient depending on the hardware. For example, the (all up, all down) encoding can be utilized not only during initialization (as discussed in Section~\ref{sec:implem} and shown in Fig.~\ref{fig:circuit}) but throughout the entire computation. In this scheme, all fermionic modes with spin up are encoded in the first $L$ qubits, while the spin down modes are encoded in the remaining qubits.

\begin{figure*}
    \centering
    \includegraphics[width=\textwidth]{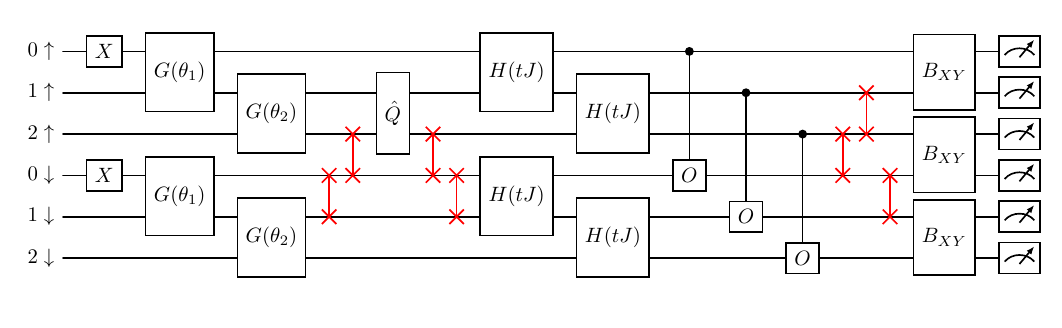}
    \caption{Complete ansatz circuit for a fermionic system of size $L=3$ in the (all up)(all down) JW ordering, where we use we encode all the spin up modes in the first $L$ qubits and the spin down modes in the last $L$.}
    \label{fig:circuit_ordering2}
\end{figure*}

Here, the initialization part of the circuit is straightforward because the Givens rotations are applied in layers independently to the spin up and spin down sector. This part of the circuit can be implemented with $L-1$ layers of Givens gates. Since every Givens rotation can be implemented with depth $2$, this part of the circuit has two-qubit gate depth $2(L-1)$.

After that, we need a FSWAP layer to bring the modes $(\lfloor L/2\rfloor,\uparrow)$ and $(\lfloor L/2 \rfloor,\downarrow)$ next to each other (both in the JW encoding and in the hardware) to implement the quench operator. Then, the FSWAP layer is applied again to go back to the original ordering. Every FSWAP layer contains $L-1$ FSWAPs. If we assume that the FSWAP gate and $\hat{Q}_{\lfloor L/2 \rfloor}=H(\pi/2)$ can be implemented with two-qubit gate depth $2$, then the part of the circuit used to implement the quench will have two-qubit gate depth $4(L-1)+2 = 4L-2$.

Similarly, assuming all-to-all connectivity, every term of the Trotterized time evolution has two-qubit gate depth $2$. Each Trotter step consists of three layers (odd hoppings, even hoppings, and on-site interactions) that can be implemented simultaneously. Therefore, the circuit for a time evolution implemented using a first-order Trotter formula with $N_{\text{Trotter}}$ Trotter steps has overall two-qubit gate depth $6N_{\text{Trotter}}$. 

Finally, we need an extra FSWAP layer (this time consisting of $(L^2-L)/2$ FSWAP gates, but still with two-qubit gate depth $2(L-1)$) to reorder the qubits so that all pairs of qubits corresponding to fermionic modes on a given site $i$ and opposite spin are next to each other. We then measure all of them in the ($XX + YY$)-basis by applying the unitary $B_{XY}$ to each pair of qubits. This is implemented with constant two-qubit gate depth $3$ and total number of two-qubit gates $3L$. 
The total two-qubit gate depth of the ansatz circuit will then be $8L+6N_{\text{Trotter}}-3$. The total circuit is shown in Fig.~\ref{fig:circuit_ordering2}.

Alternatively, we can leave out the last FSWAP layer at the price of increasing the number of circuits we need to run. That is, for every observable $\hat{S}^x_i$ with $i\in\{0,\dots, L-1\}$, we can consider a circuit where we measure the high-weight Pauli strings $\braket{X_i Z_{i+1} \dots Z_{i+L-1} X_{i+L}}$ and $\braket{Y_i Z_{i+1} \dots Z_{i+L-1} Y_{i+L}}$, and add them together. Note that this strings arise as a consequence of JW encoding when two fermionic modes are not adjacent to each other. In this case, the depth will be $6L+6N_{\text{Trotter}}-4$ instead. However, the sampling overhead for this approach grows with the system size and the expectation values $\braket{\hat{S}^x_i}$ are more sensitive to noise. Moreover, since in this case one has to measure the two parts of $\braket{\hat{S}^x_i}$ independently, post-selection on fermionic occupation number is not possible. For all these reasons, we have not considered this approach in this work.

Note that, using the DGA scheme to prepare an approximate free-fermion ground state, the overall two-qubit gate depths become $4n_{\mathrm{layers}}+6L+6N_{\text{Trotter}}-1 $ and $4n_{\mathrm{layers}}+4L+6N_{\text{Trotter}}-2$ with and without the last FSWAP layer, respectively.

Comparing these resource requirements with the one discussed in Section~\ref{sec:implem}, we can see that the (all up)(all down) encoding is more efficient in the presence of small system size and large number of Trotter steps. For the values of $L$ and $N_{\mathrm{Trotter}}$ considered in this work (see Table~\ref{tab:exp_params}), we have found that the (up, down)(up, down) JW ordering is always the most convenient. Moreover, one has to keep in mind that in order to implement the time-evolution operator associated with $H_{\rm int}$ on a device with local connectivity, as the one we consider in this work, the interacting qubits must be physically close to each other. In the (all up)(all down) encoding, this requires additional SWAP gates. On the other hand, the implementation in the (up, down)(up, down) encoding is straightforward.

\section{Resilience to particle number errors}
\label{app:depolarizing}

\begin{figure*}
    \centering
    \includegraphics[width=0.9\linewidth]{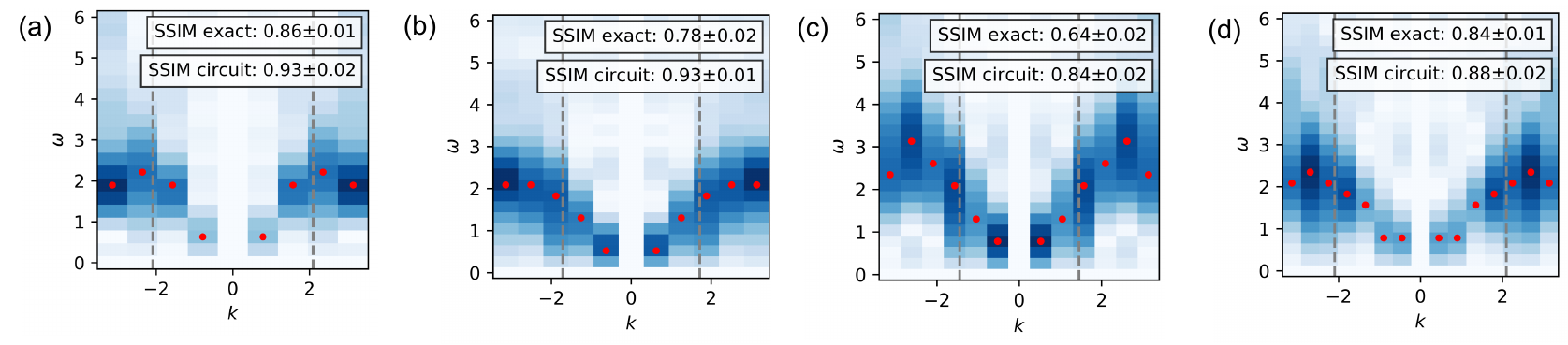}
    \caption{Hardware results from quench spectroscopy simulations of the 1D Fermi-Hubbard model for system sizes $L=9, 11, 13, 15$ (from left to right) without particle number post-selection. The circuit details and the parameters are the same as in Figure \ref{fig:real_hardware} and
    reported in Tables~\ref{tab:resources} and \ref{tab:exp_params}.
    }
    \label{fig:real_hardware_no_selection}
\end{figure*}

As we observed in Section~\ref{sec:hardware_results}, the QSF that we extract in the quench spectroscopy experiment is somewhat resilient to bit-flip errors. As an example, in Fig.~\ref{fig:real_hardware_no_selection} we show the hardware results of panels (e-h) in Fig.~\ref{fig:real_hardware} without particle number post-selection. Although the quality of the reconstructed spectra is slightly reduced, all the main features can still be observed. 
In the following, we discuss a simplified model for these errors that explains this property. We consider the following bit-flip channel
\begin{align}\label{eq:bit_flip}
    \Omega(\rho):=\left(1-\sum_i\epsilon_i\right)\rho+\sum_i\epsilon_i(X_i\rho X_i),
\end{align}
where $\epsilon_i$ is the probability of qubit $i$ flipping. We assume a simple scenario where the spin-flips are just related to measurement, so they happen at the end of the circuit. In this case, the full density matrix becomes $\Omega(\rho(t))$, where $\rho(t)$ is the initial density matrix (see, e.g., Eq.~\eqref{eq:exp_2}) evolved with the time-evolution unitary operator $e^{-iHt}$. The expectation value of the spin operator $\hat{S}_j^x$ on this noisy density matrix is
\begin{align}
    {\rm Tr}\left(\hat{S}_{j}^{x}\Omega(\rho(t))\right)&=\left(1-\sum_{i}\epsilon_{i}\right){\rm Tr}\left(\hat{S}_{j}^{x}\rho(t)\right) \nonumber\\
    &+\sum_{i}\epsilon_{i}{\rm Tr}\left(X_{i}\hat{S}_{j}^{x}X_{i}\rho(t)\right).
\end{align}
We can analyse the last contribution in the equation above by expressing the fermionic operator $\hat{S}_j^x$ in terms of Pauli operators, using the JW encoding ordering discussed in Section \ref{sec:implem}, as follows
\begin{align}\nonumber
    {\rm Tr}\left(X_{i}\hat{S}_{j}^{x}X_{i}\rho(t)\right)&=\frac{1}{2}{\rm Tr}\left[X_{i}\left(X_{j\uparrow}X_{j\downarrow}+Y_{j\uparrow}Y_{j\downarrow}\right)X_{i}\rho(t)\right]. 
\end{align}
Note that due to the particular JW ordering of the measurement, where the modes corresponding to fermions on the same site and opposite spin are next to each other, $\hat{S}_j^x$ does not have a string of $Z$ operators so that
\begin{align}\nonumber
X_{i}\hat{S}_{j}^{x}X_{i}&=\delta_{i\neq(j\uparrow / j\downarrow)}\left(X_{j\uparrow}X_{j\downarrow}+Y_{j\uparrow}Y_{j\downarrow}\right)\\&+\delta_{i=j\uparrow / j\downarrow}(X_{j\uparrow}X_{j\downarrow}-Y_{j\uparrow}Y_{j\downarrow}).
\end{align}
In the first term of the equation above we recognize the spin operator $\hat{S}_j^x$, while the second one corresponds in the fermion basis to
\begin{align}
    \frac{1}{2}\left(X_{j}X_{j+1}-Y_{i}Y_{j+1}\right)=-(c_{j\uparrow}c_{j\downarrow}+c_{j\downarrow}^{\dagger}c_{j\uparrow}^{\dagger}).
\end{align}
The trace of this last term on the state $\rho(t)$ is zero, as the state has a definite number of particles. This means that the original signal gets damped by a factor that does not depend on the size of the system
\begin{align}\label{eq:error_chan}
    {\rm Tr}\left(\hat{S}_{j}^{x}\Omega(\rho(t))\right)=(1-\epsilon_j)
    {\rm Tr}\left(\hat{S}_{j}^{x}\rho(t)\right).
\end{align}
As we discussed in Section~\ref{sec:hardware_results}, assuming $\epsilon_j\approx\epsilon$ $\forall j$, such a damping does not affect the reconstructed QSF.
Defining $\rho_j(t):=e^{-iHt}|\Psi^j_{\frac{\pi}{4}}\rangle\langle\Psi^j_{\frac{\pi}{4}}|e^{iHt}$, and
\begin{align}
\mathcal{S}_q[\rho_j](\omega):=\sum_{l=0}^{L-1}\int_0^\infty dt\,e^{i\omega t}e^{iql}{\rm Tr}(\hat{S}_{j+l}^{x}\rho_j(t)).
\end{align}
The previous analysis can be expressed as the following lemma:\\
\textbf{Lemma:} The action of a bit-flip channel $\Omega$ defined in Eq.~\eqref{eq:bit_flip} with $\epsilon_i=\epsilon$, $\forall i$, on the time-evolved density matrix produces the following multiplicative error in the QSF,
\begin{align}
    |\mathcal{S}_q[\rho_j](\omega)-\mathcal{S}_q[\Omega(\rho_j)](\omega)|\leq \epsilon |\mathcal{S}_q[\rho_j](\omega)|.
\end{align}
The proof follows directly from the definitions and Eq.~\eqref{eq:error_chan}.
This result generalizes straightforwardly to the case of a fully depolarizing channel.

\section{Resilience of spectroscopic signals from perturbation theory}
\label{app:robustness}

In this section, we use perturbation theory to demonstrate the robustness of the quench spectroscopy protocol. As we will show, the shape of the Fourier transform of the retarded spin Green's function is heavily influenced by the phase space available for the excitations. The latter is resilient to small imperfections in the Hamiltonian and in the initial state. Therefore, in the weak interaction regime, the error introduced in the quench spectral function by starting the quench spectroscopy protocol from the free-fermion model ground state is perturbatively small.

For simplicity, let us assume that the system has full translational invariance, so Bloch-momentum is a good quantum number.
We define the spin operator in momentum space as
\begin{align}
    \tilde{S}_{k}^{x}&:=\frac{1}{\sqrt{L}}\sum_{j=0}^{L-1} e^{ikj}\hat{S}^x_{j},
\end{align}
where $L$ is the number of sites and $k= \pi n/L$ with $n\in \mathbb Z_L$.
The time ordered spin Green's function in momentum-frequency space is in turn defined as 
\begin{equation}
    \mathcal{G}(k,\omega):=\int_{-\infty}^{\infty}e^{i\omega t}i\langle\Psi|\mathcal{T}(\tilde{S}_{k}^{x}(t)\tilde{S}_{-k}^{x})|\Psi\rangle dt,
\end{equation} 
where $\mathcal{T}$ is the time-ordering operator \cite{fetter2003quantum}.
The equivalent expression in the Lehmann representation of the time-ordered spin Green's function is
\begin{align}
    \label{eq:Imag_spin_GF}
    \mathcal{G}(k,\omega)&=\sum_{m}\frac{|\langle\Psi|\tilde{S}_{-k}^{x}|E_{m}\rangle|^{2}}{\omega-(E_{m}-E_{0})-i\eta}
    -\frac{|\langle\Psi|\tilde{S}_{k}^{x}|E_{m}\rangle|^{2}}{\omega+E_{m}-E_{0}+i\eta},
\end{align}
where $E_m$ are the eigenenergies of the Hamiltonian associated with eigenstates $\ket{E_m}$ and $\eta\rightarrow 0^+$ is a normalization parameter to control the convergence of the time integral.

\begin{figure}[t]
    \centering
    \includegraphics[width=0.35\textwidth]{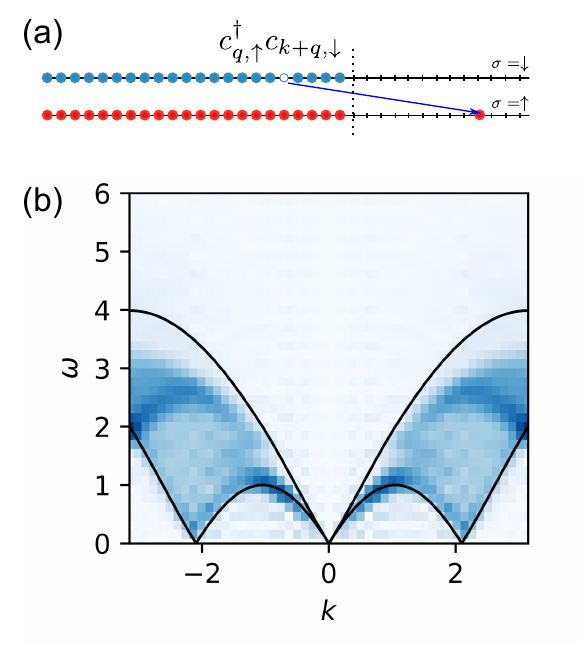}
    \caption{(a) Diagram of creation of a spinon. Spin excitations can be created from the filled Fermi surface (with Fermi momentum $k_F$ denoted by the dashed line) by removing an electron from a spin sector within the Fermi surface and creating an electron outside it with different spin. (b) Comparison between the boundaries of spin excitation spectrum for an non-interacting system of infinite size (black contour) and a system with $U=3$ and $L=51$ (same as Fig.~\ref{fig:classical_simulation}(b)). The value on the horizontal axis corresponds to the total momentum of the pair of spinons $k$.}
    \label{fig:comp_u_zero}
\end{figure}

The spin Green's function satisfies the Dyson equation \cite{fetter2003quantum},
\begin{equation}
    \mathcal{G}(k,\omega)=\frac{1}{[\mathcal{G}^0(k,\omega)]^{-1}-\Sigma^*(k,\omega)}.
\end{equation}
Here, $\Sigma^*(k,\omega)$ is the self-energy for the spin  Green's function and $\mathcal{G}^0(k,\omega)$ is the spin  Green's function for the non-interacting system. Clearly, in the absence of interactions, $\Sigma^*(k,\omega)=0$. 
Let us now focus on the weak interaction regime $U\ll J$. In this case, the interacting spin GF is mainly determined by the non-interacting one, except where $[\mathcal{G}^0(k,\omega)]^{-1}=0$ as in this region the interacting spin Green's function is {\it solely} determined by interactions. We can use Eq.~\eqref{eq:Imag_spin_GF} to determine the region $[\mathcal{G}^0(k,\omega)]^{-1}=0$ which is the region most strongly affected by interactions, even in the perturbative regime. To this end, we have to understand the matrix element $A_{m,k}:=\bra{E_m}\tilde{S}_k^x\ket{\Psi_0}$ where $\ket{\Psi_0}$ is the ground state of the non-interacting system. Writing explicitly the spin operator in terms of the creation and annihilation operators in momentum space we have
\begin{align}
    \label{eq:spin_op_momentum} 
    \tilde{S}_{k}^{x}&=\frac{1}{\sqrt{L}}\sum_{j}e^{ikj}\frac{1}{L}\sum_{q_{1}q_{2}}e^{i(q_{1}-q_{2})j}\left(c_{q_{1}\uparrow}^{\dagger}c_{q_{2}\downarrow}+c_{q_{1}\downarrow}^{\dagger}c_{q_{2}\uparrow}\right) \nonumber\\
    &=\frac{1}{\sqrt{L}}\sum_{q}\left(c_{q,\uparrow}^{\dagger}c_{k+q,\downarrow}+c_{q,\downarrow}^{\dagger}c_{k+q,\uparrow}\right).
\end{align}
The presence of the Fermi surface constraints the allowed values of $k$ and $q$ where $A_{m,k}$ is nonzero. The operator $c_{q,\downarrow/\uparrow}$ can only act non-trivially on occupied states in the Fermi sea, so $|q|\leq k_F$, while the only states to be excited by $c^\dagger_{q,\uparrow/\downarrow}$ are the states outside the Fermi sea, i.e., $k_F\leq |q|\leq \pi$. A sketch of these processes is shown in Fig.~\ref{fig:comp_u_zero}(a). Then the region $[\mathcal{G}^0(k,\omega)]^{-1}=0$ corresponds to all $q$ such that for a given $k,\omega$
\begin{align}\nonumber
    \omega -(E_m-E_0)&=\omega + 2J[\cos(q)-\cos(k+q)]=0.
\end{align}
Defining $q=\frac{k+\delta q}{2}$, the set of frequencies where $[\mathcal{G}^0(k,\omega)]^{-1}=0$ becomes simply
\begin{align}
    \omega = 4J \sin\left(\frac{k}{2}\right)\sin\left(\frac{\delta q}{2}\right).
\end{align}
Plotting this relation as a function of $k$ for all the allowed values of the momentum difference $\delta q$, i.e., $2k_F-k\leq \delta q\leq 2\pi -k$ and $-2\pi-k\leq \delta q\leq -2k_F -k$, leads to a continuum of states whose boundaries are shown as black lines in Fig.~\ref{fig:comp_u_zero}(b). 

This shows that the effect of interactions is to renormalize the boundaries of the region defined by $[\mathcal{G}^0(k,\omega)]^{-1}=0$, which is primarily defined by the phase space available for the excitations in presence of a Fermi sea. 

Furthermore, in the weak interaction regime, the fact that the boundaries of the spectral signal are mainly determined by the phase space available for the excitations continues to hold even in the presence of errors in the implementation of the dynamics of the system, such as the ones arising due to Trotterization and/or hardware noise. To see this, let us assume that, as shown in Appendix~\ref{app:depolarizing} for the case of a bit-flip channel, the error in the spin Green's function is multiplicative. We can then relate the experimental signal $\mathcal{\tilde{G}}(k,\omega)$ to the exact signal as
\begin{align}\nonumber
    \mathcal{\tilde{G}}(k,\omega)&=(1+\epsilon(k,\omega))\mathcal{G}(k,\omega)\\
    &=\frac{1+\epsilon(k,\omega)}{[\mathcal{G}_{0}(k,\omega)]^{-1}-\Sigma^*(k,\omega)}
\end{align}
where $\epsilon(k,\omega)$ characterizes the error, which we assume to be small, i.e., $|\epsilon(k,\omega)|\ll 1$. Although a full characterization of this error function is beyond the scope of this work, we assume a dependence on the frequency and momentum to account for imperfections in the time evolution, in the circuit structure, and hardware noise. It is clear that the previous analysis of the phase space dominating the spectral signal for weak interactions remains valid.

\section{Analysis of the DGA scheme}
\label{app:DGA}

As discussed in the main text, the DGA algorithm allows us to prepare an approximate free-fermion initial state using a shallow circuit consisting of the Givens rotation layers shown in Fig.~\ref{fig:schematic_approx_givens_layer} whose angles are optimized using VQE and the fact that true free-fermion ground state is classically efficient to compute~\cite{Terhal_2002}. 

\begin{figure}
    \centering
    \includegraphics[width=0.49\textwidth]{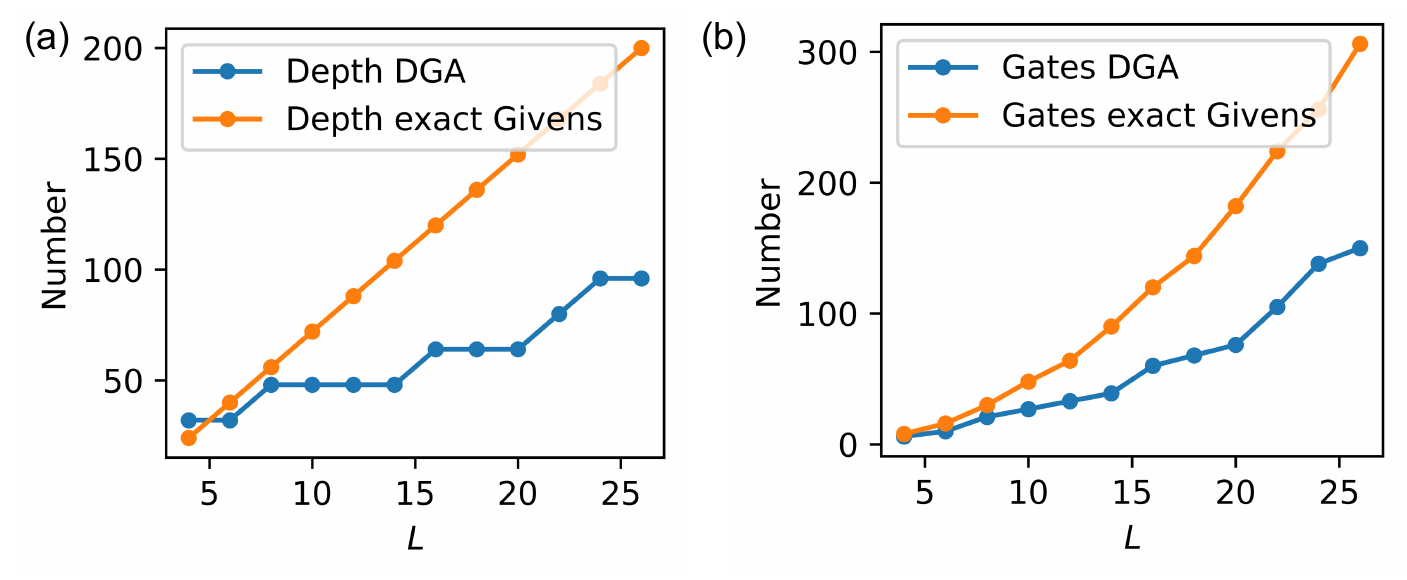}
    \caption{Total two-qubit gate circuit depth (a) and gate count (b) required for preparing the exact free-fermion ground state using Givens rotations (blue curve) compared with the ones required by the DGA scheme (yellow curve) as a function of the system sizes at filling $n_e = 2/3$. The VQE optimization of the DGA Givens angles is done through minimizing the energy and the fidelity is asserted to be $|\braket{\psi_{\rm FF}|\psi_{\rm DGA}}|^2 \geq 0.95$. The numbers here are quoted for combining both spin sectors.}
    \label{fig:approx_ff_layers_wrt_size}
\end{figure}

For concreteness, in Fig.~\ref{fig:approx_ff_layers_wrt_size} we compare the two-qubit gate depth and count required to prepare the free-fermion ground state using the QR decomposition~\cite{jiang2018quantum}, $\ket{\psi_{\rm FF}}$, and the approximate DGA state, $\ket{\psi_{\rm DGA}}$. Here, we fix the electronic filling as $n_e=2/3$ for various 1D system sizes and assert that the fidelity overlap of the DGA state and the exact free-fermion ground state is set at $|\braket{\psi_{\rm FF}|\psi_{\rm DGA}}|^2\geq 0.95$. From this study, it is clear that despite the number of Givens rotation layers needed to achieve an approximation of the free-fermion ground state with a certain fidelity is $O(L)$, as in the case of the QR decomposition, the smaller prefactor of the DGA algorithm allows us to attain a significant saving in circuit depth and gate counts already at very small system sizes. 

\begin{figure}
    \centering
    \includegraphics[width=0.35\textwidth]{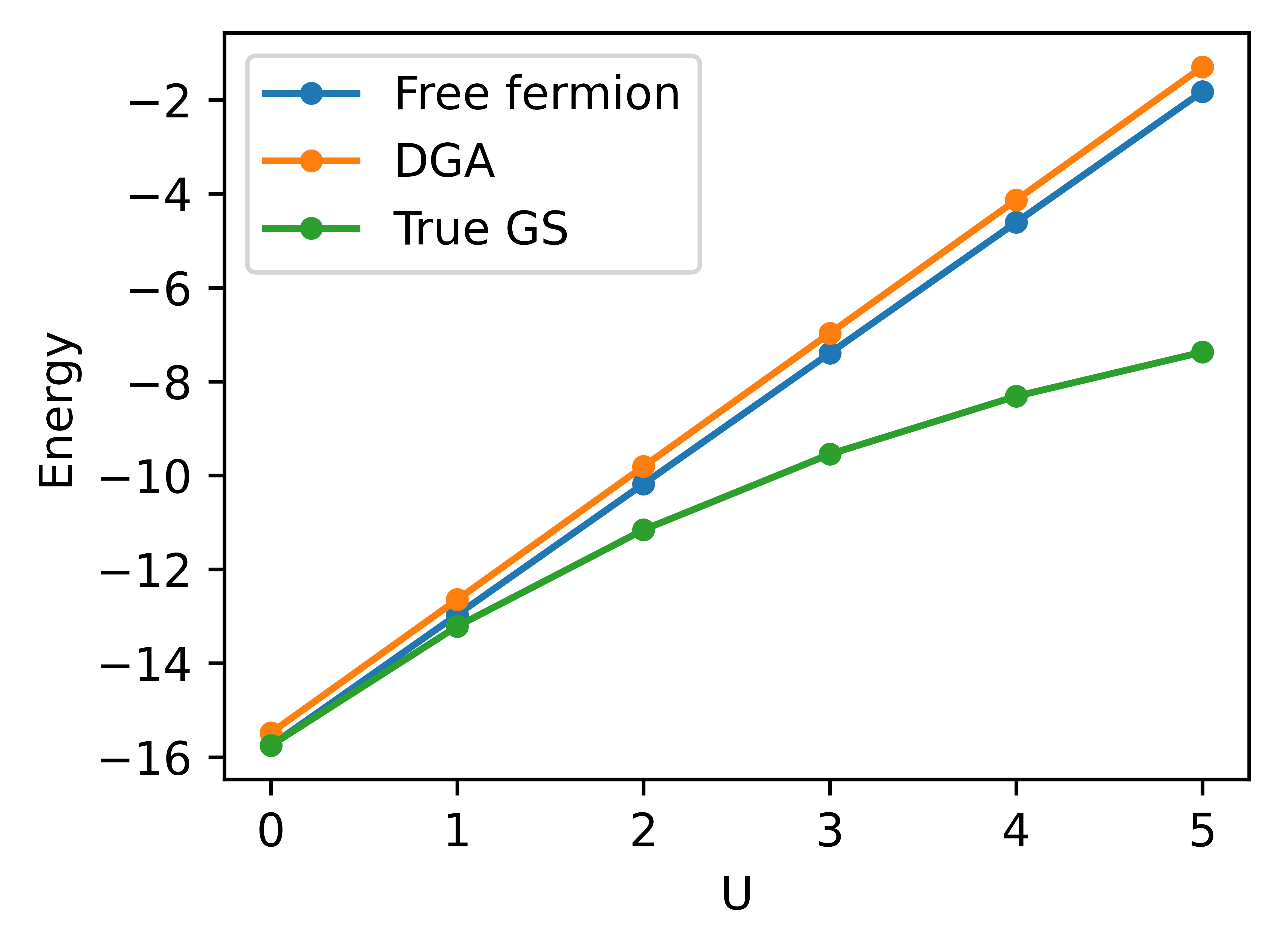}
    \caption{Comparisons of the ground state energy between the true interacting ground state (green curve) at various interaction strengths $U$ against the exact free-fermion state (blue curve) and approximate free-fermion state obtained using 2 layers of the DGA scheme (yellow curve) for $L=13$ with $N_e=6$ electrons.
    }
    \label{fig:ff_vs_approx_vs_trueGS}
\end{figure}

For completeness, we also show in Fig.~\ref{fig:ff_vs_approx_vs_trueGS} the same energy analysis as that in Fig.~\ref{fig:free_fermion_simulation}(a) in the main text by including the energies of the approximate free-fermion ground state obtained via DGA for $L=13$. We numerically estimate that the energy injected by the quench of Eq.~\eqref{eq:quench_op_unitary} is $\Delta E\approx 0.6$ (see Fig.~\ref{fig:free_fermion_simulation}(a)). The energy difference between the two states is negligible compared to this energy scale and we can therefore expect that using the DGA approximate state will not affect the performances of the quench spectroscopy protocol.

\begin{figure}
    \centering
    \includegraphics[width=0.35\textwidth]{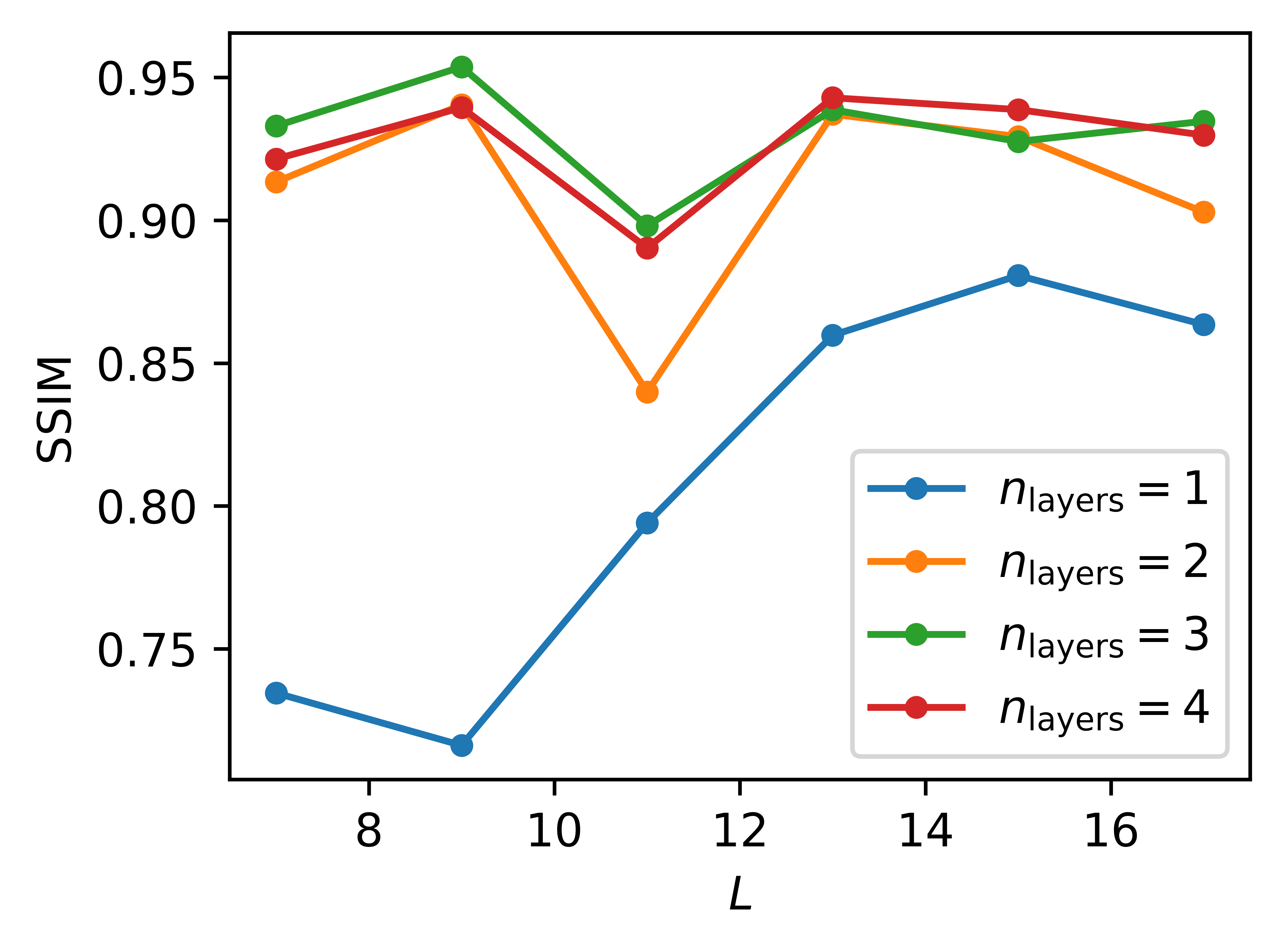}
    \caption{Structural similarity index measure (SSIM) of the spin excitation spectrum obtained with the approximate DGA state with respect to one computed using the exact ground state for different numbers of DGA layers $n_{\rm layers}$. The SSIM is shown as a function of the system size $L$ for a fermion density close to $n_e=2/3$.}
    \label{fig:SSIM_DGA}
\end{figure}

In Fig.~\ref{fig:approx_ff} in the main text, we demonstrate how the fidelity depends on the system size and the number of DGA layers. We observe a consistent increase in fidelity with the addition of more layers for all system sizes $L$. To ensure that this behavior extends to the spectrum—the final quantity of interest in this work—we compared the exact spectrum with the one obtained by initializing the protocol using DGA with $n_{\text{layers}}$ layers. In Fig.~\ref{fig:SSIM_DGA}, we present the SSIM of the spectrum generated by evolving the state for a total time $T=3$ using $N_{\text{trotter}}=5$ Trotter steps. 

As anticipated, the SSIM improves with an increasing number of layers, surpassing the 0.9 threshold for $n_{\rm layers}>3$ DGA layers, at which point the differences become indistinguishable. Also, notice that the decay of the SSIM as a function of $L$ for $n_{\rm layers}>3$ is barely noticeable. Hence, we expect that shallow DGA states will allow one to recover excitation spectra even in larger systems.

\input{main_arxiv.bbl}

\end{document}

%% file: main_arxiv.bbl
%